\documentclass[fleqn,10pt,twocolumn]{wlscirep}
\usepackage[utf8]{inputenc}
\usepackage[T1]{fontenc}
\usepackage[left]{lineno}
\graphicspath{{figures/}}
\usepackage{subfigure}
\usepackage{braket}
\usepackage{pdflscape}
\begin{document}

\title{\Large Real-time identification of the onset of financial rogue waves}

\author[1,*]{\small Rosie Hayward}
\author[2]{\small Orla Lennon}
\author[2]{\small Fabio Biancalana}
\affil[1]{Supply Chain Intelligence Institute Austria, Vienna, Austria}
\affil[2]{Heriot-Watt University, Edinburgh, United Kingdom}

\affil[*]{hayward@ascii.ac.at}

\begin{abstract}
Extreme events in financial systems, often captured by indicators such as volatility, remain difficult to identify close to their onset. Volatility shares many statistical properties with other natural, complex systems which experience extreme events, which we explore in this manuscript. We extend the analogy between rogue waves in optical and hydrodynamical systems to financial volatility by identifying rogue-wave-like peaks with similar statistical properties. We use a Schr\"odinger equation where the potential follows the shape of a Kerr nonlinearity to examine the properties of financial volatility indices within a moving time window. We see evidence of Anderson localisation as a rogue peak approaches in the VIX, and show that the numerical gradient of the system's minimum eigenvalue reliably spikes at the onset of an extreme event. We adapt our methodology to simulate the real-time arrival of data, and show that all but one of the VIX's major peaks can be detected given a reasonable amount of history. We then perform two out-of-sample tests, one for the VXO index, and one for the VSTOXX index, and successfully replicate our initial results, identifying all but one major peak (87.5\% or 7/8) in both cases. This method of analysis shows considerable promise as a tool for identifying potential financial crises, aiding in their mitigation.
\end{abstract}

\maketitle

In this work, we address the question of whether the onset of extreme financial events in volatility indices can be further understood and even detected, by harnessing analogies with other complex systems. The universality of stylised facts in complex systems, particularly the university of power laws or fat-tailed distributions, and the mechanisms which could lead to them such as self-organised criticality, hidden variables, preferential attachment, and multiplicative processes more generally, have been well explored in the literature~\cite{guerrero_multiplicative_2020}. This has led to complex systems approaches to asset bubble detection and prediction, perhaps most notably the Log Periodic Power Law model~\cite{sornette_stock_1996}, with recent research combining these models with the use of AI achieving a maximum precision of 50.88 when identifying crashes\%~\cite{lee_more_2025}. Exploring events with similar properties observed in other complex systems may then offer an opportunity for improved detection.

Rogue waves are an example of an extreme event which emerges from a natural, complex system~\cite{dudley_rogue_2019}. The first measured rogue wave was the oceanic Draupner wave, recorded in the North Sea in 1995 and reaching 25.6m in height and 18.5m in elevation~\cite{haver_possible_2004}. This event arguably spurred scientific interest in rogue waves, with an analogy between extreme intensities found in optical fibres and the generation of large ocean waves drawn in 2007 leading to many additional avenues of investigation~\cite{solli_optical_2008,dudley_rogue_2019}. 

The long (or fat)-tailed distributions which motivated the original analogy between optical and oceanic waves are a feature of many other systems displaying extreme events, such as the Gutenburg-Richte law for the intensity of earthquakes~\cite{gutenberg_frequency_1944}, the distribution of wealth, rainfall depth, city size, as well as financial asset returns~\cite{hutt_synergetics_2020} and the volatility of price fluctuations~\cite{liu_statistical_1999}. Analogies between extreme events in finance and other complex systems are well covered in the literature, with the Omori power-law relaxation following an earthquake~\cite{parsons_global_2002} also seen in limit order books and volatility indices ~\cite{toth_studies_2009,lillo_power-law_2003}. This analogy extends to fluid flow, with volatility clustering seen in price movements mirroring the intermittency seen in fluid turbulence~\cite{pasquini_multiscale_1999}. The long-tailed behaviour of asset returns also mimics the distribution of flow in a turbulent fluid: when comparing the distribution of sizes of fluctuations, changing the time scale at which fluctuations are compared can result in a divergence from a normal distribution and the emergence of fat tails~\cite{ghashghaie_turbulent_1996,hutt_synergetics_2020}. 

The view of markets as complex, adaptive systems which could display nonlinear behaviour contributed to the proposal of a nonlinear option pricing model and a subsequent paper describing the existence of financial rogue waves in this model \cite{ivancevic_adaptive-wave_2010,yan_financial_2010}. The authors of this manuscript do not suggest that quantum mechanics is applicable to financial systems, however analogies in science can be useful and powerful tools for analysis. Here, we ask, given the many analogous behaviours between systems featuring extreme events and cascades~\cite{aubrun_identifying_2025}, whether the science of rogue waves in the classical systems of optics and hydrodynamics could aid us in understanding the onset of financial extreme events. 

An ongoing question in the field of rogue waves is how linear and nonlinear focusing may interact in their generation. Both linear and nonlinear focusing can contribute to the generation of optical rogue waves and rogue waves in water tanks, however the importance of nonlinearity in oceanic rogue waves is debated~\cite{dudley_rogue_2019}. A key component of this discussion is the nonlinear Schr\"odinger equation (NLSE). This equation describes slowly modulated wave groups on the surface of a deep fluid~\cite{zakharov_stability_1972}, as well as the slowly varying envelope of an optical field in a nonlinear (Kerr) medium~\cite{chiao_self-trapping_1964}, and is known as universal in its description of slowly-varying, quasi-monochromatic wave packets in weakly nonlinear media~\cite{kato_nonlinear_2005}. Saleh et. al. demonstrated in 2017 that when the refractive index is modulated by a Kerr nonlinearity, the modes of the Schr\"odinger equation generated through Anderson localisation can experience solitonisation ~\cite{saleh_anderson_2017}. In water, nonlinearity was shown to strengthen the localisation of surface gravity waves propagating in a canal with a random bottom~\cite{ricard_effects_2024}. We consider this interaction between linear and nonlinear effects to be of potential importance in the generation of extreme financial events, which motivates our methodology.

In analogy with optical systems, using the Schr\"odinger equation where the potential takes the shape of a Kerr nonlinearity~\cite{saleh_anderson_2017}, we look for evidence of Anderson localisation in the run-up to extreme financial events in three volatility indices: the Cboe VIX\textsuperscript{\tiny\textregistered} index~\cite{noauthor_volatility_2024}, which estimates the expected volatility of the S\&P 500\textsuperscript{\tiny\textregistered}  index using option prices, the VXO\textsuperscript{\tiny\textregistered} index, which corresponds to the former methodology of the VIX and estimates the expected volatility of the S\&P 100\textsuperscript{\tiny\textregistered}  index, and the VSTOXX\textsuperscript{\tiny\textregistered}  index, which estimates the expected volatility of the EURO STOXX 50\textsuperscript{\tiny\textregistered}  index, also using option prices\cite{seegopaul_vstoxx_2024}. We chose to examine financial volatility due to its lack of long-term trend, easily identifiable extreme peaks, many of which correspond to large financial crises, and its long-tailed nature. By extracting the envelope wave of each index, we can aim to remove small fluctuations which do not contribute to the long-run behaviour of the system (either small exogenous shocks from which the system recovers quickly \cite{aubrun_identifying_2025}, or endogenous behaviour which happens on too short a timescale to resolve) and identify changes in the behaviour of the system as an extreme event approaches. 

Here, we show how use of the rogue-wave analogy allows extreme events in these indices to be easily identified, and recover the long-tailed statistics and power-law relaxation of the highs of the VIX. We then demonstrate how when taking the squared-envelope function of the VIX within a moving window as a potential in the Schr\"odinger equation, the numerical gradient of the absolute value of the minimum eigenvalue of the system systematically spikes before an oncoming extreme event. We show how using this numerical eigenvalue gradient a reliable indicator of an oncoming extreme event can be obtained for the VIX index when simulating real-time receipt and analysis of the data to prevent information leakage. We then provide the results of two out-of-sample tests, one for the highs of the VXO and one for the highs of the VSTOXX, where we show thresholds estimated based on the indicator values obtained during in-sample testing are successful in identifying the onset of all but one of the most extreme peaks in both cases. 

While many methods of volatility forecasting exist, they typically do not seek to identify the onset of extreme events, instead focusing on the value-at-risk or expected shortfall within a given period of time~\cite{hutt_synergetics_2020}. The methodology we present is not only novel, but unique in its ability to reliably identify changes in volatility which precede a cascade of events leading to an extreme peak. This suggests a fundamental difference between spikes in volatility which precede an extreme event, and those which are simply noise, opening up the possibility of greater understanding of the underlying dynamics which drive a large-scale crash in prices.

\section*{Results}

\subsection*{Identifying rogue waves in the VIX}

Several criteria have been identified as important in the classification of rogue waves~\cite{akhmediev_editorial_2010}, two of which can be applied directly to the VIX. Firstly, the height of the wave, the distance from the wave trough (lowest point) to the wave crest (highest point), should exceed 2 to 2.5 times the significant wave height -- traditionally defined as the average of the top third of wave heights~\cite{dysthe_oceanic_2008}. Secondly, they appear more often than Gaussian statistics would suggest, following a long-tailed distribution and displaying a power-law relationship between occurrence and height, seen both in ocean physics and in optical media~\cite{dudley_rogue_2019,cattrell_can_2018,solli_optical_2008}. 

\begin{figure}[ht]
\centering
   \includegraphics[width=1\linewidth]{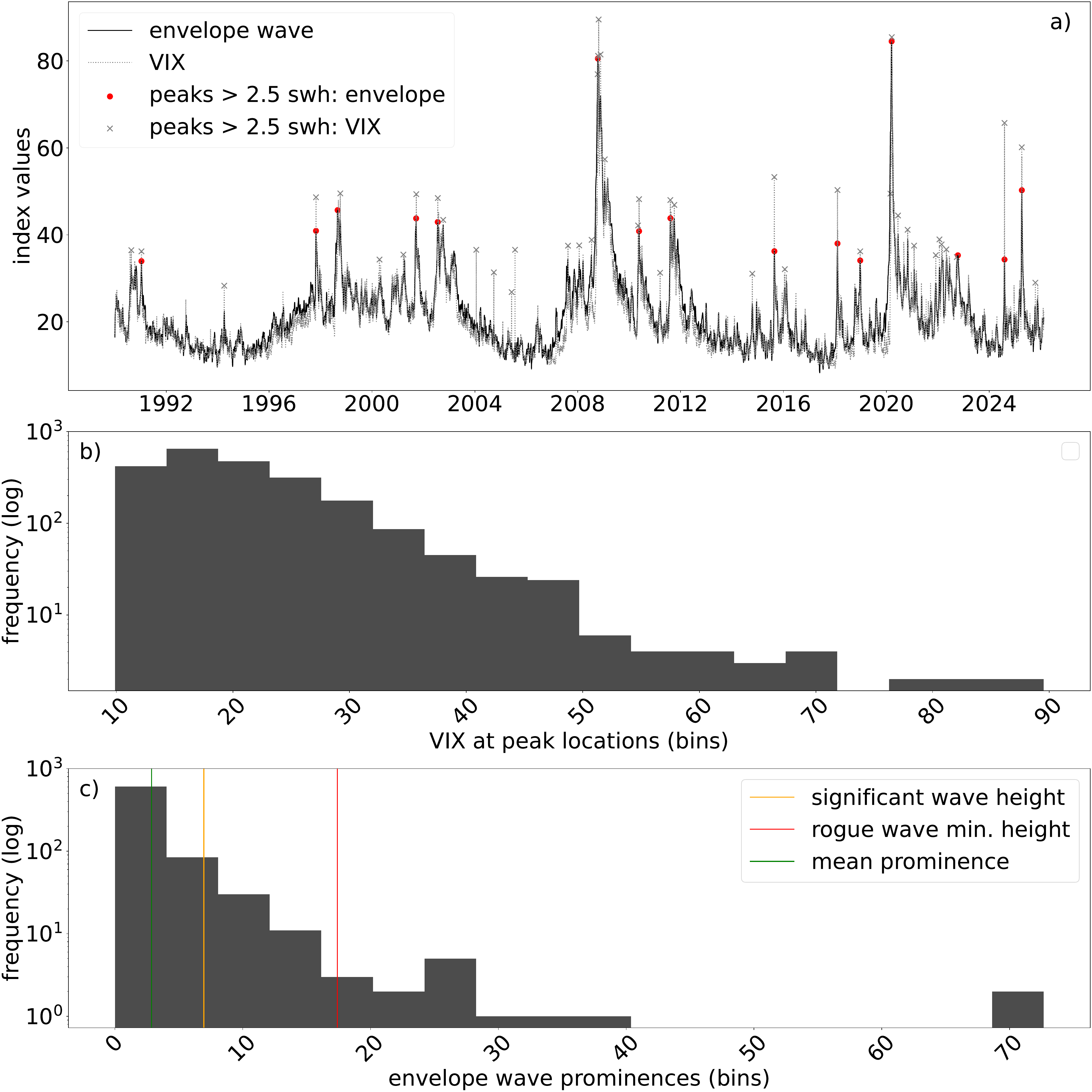}   
    \caption{a) VIX highs (light grey dashed) with extreme event peaks ($>$ 2.5 SWH) marked by grey crosses, envelope wave (black), with extreme event peaks marked by red dots ($>$ 2.5 SWH) and red crosses ($>$ 2 SWH). A horizontal line marks where the VIX exceeds 35. b) Frequency distribution of raw heights (values at the peak locations) of the VIX highs, with a vertical orange line marking a height of 35. c) Distribution of the prominences of the peaks found for the envelope wave.} \label{fig:peaks}
\end{figure}

In order to identify rogue wave-like extreme events in the VIX and exclude fluctuations which happen over small timescales, we process the time series to extract the envelope wave of the signal, discarding any peaks below the 92.5th percentile in Fourier space. We then use a SciPy's find\_peaks function to identify peaks, highlighting those which exceed both 2 and 2.5 times the significant wave height (SWH) in both the raw and processed VIX data, based on the average prominence of peaks in the entire time series. For this application, we assume peak prominences are suitably analogous to the height of ocean waves (peak to trough) typically used when calculating the significant wave height, particularly as we are interested in the wave's envelope. The raw VIX and corresponding envelope wave with extreme peaks identified can be seen in figure~\ref{fig:peaks}. Here, the raw VIX is displayed in dashed grey, and the envelope wave with small fluctuations removed in black. The peaks identified as rogue waves ($>$ 2.5 SWH) for the unprocessed time series are marked by grey crosses, and the peaks identified as rogue waves in the processed time series are marked with red dots ($>$ 2.5 SWH) and red crosses ($>$ 2 SWH) . A grey horizontal line marks where the VIX index is equal to 35 -- a threshold which many larger financial crashes exceed and which reflects large expected price movements by investors. 

The removal of small fluctuations in the signal produces the expected outcome of the peak-finding function identifying fewer peaks overall and fewer extreme peaks, which can be seen to correspond to larger movements in the envelope function of the VIX rather than sudden spikes. This can be seen by comparing the peaks marked by grey crosses with the peaks marked by red dots or crosses in the top panel of figure \ref{fig:peaks}. The frequency distribution of the raw values of the VIX at all peak locations can be seen in figure \ref{fig:peaks} b), with the orange line marking the VIX value of 35. The distribution of the prominences (the vertical distance between the peak and its lowest contour line) of the peaks found for the processed signal can be seen in figure \ref{fig:peaks} c). Here, the green vertical line marks the mean prominence of all identified peaks, the orange line marks the significant wave height, and the red line marks the minimum height of a rogue wave or extreme event. These frequency distributions are in line with the known long-tailed nature of volatility distributions, and illustrate an interesting decay relationship between the occurrence of the envelope wave's peaks and their prominences.

We include further results in the supplementary material which support this analogy. Supplementary figure S1 demonstrates how the SWH threshold can be used to count aftershocks in volatility which follow an Omori-like law, similar to what is seen in~\cite{lillo_power-law_2003} for the cumulative number of times returns of the S\&P 500 exceeded a fixed threshold, demonstrating power law relaxation. Supplementary figure S2 further demonstrates the appearance of long tails when plotting the probability distribution of differences in the values of an envelope wave which occur a certain time-frame apart, analogous to what is seen for turbulent flow in fluids~\cite{ghashghaie_turbulent_1996, hutt_synergetics_2020}.

These collective results, while also seen in previous works on financial time series, allow us to confirm that the peaks identified according to our novel application of the significant wave height criteria to the VIX meet the general definition of rogue waves, and furthermore behave similarly to waves in nonlinear water and optical systems. This analogy is key, and it is central to the motivation for our methodology in identifying the onset of extreme events.

\subsection*{Warning signals before extreme events}
In a previous study by Saleh et. al., the Anderson localisation of light was explored as a potential mechanism which contributes to the creation of optical rogue waves. To study this phenomenon, the authors examined the instantaneous, linear states of a system described by the NLSE due to the modulation of the refractive index of the medium. We explore the possibility of Anderson localisation in the VIX index before an extreme event by drawing an analogy with this methodology, as outlined in methods section, through analysing the eigenstates and eigenvalues of the potential (equation \ref{eq:potential}) for a given window length L, as well as its skewness and kurtosis.
\begin{figure}[ht]
\centering
   \includegraphics[width=\linewidth]{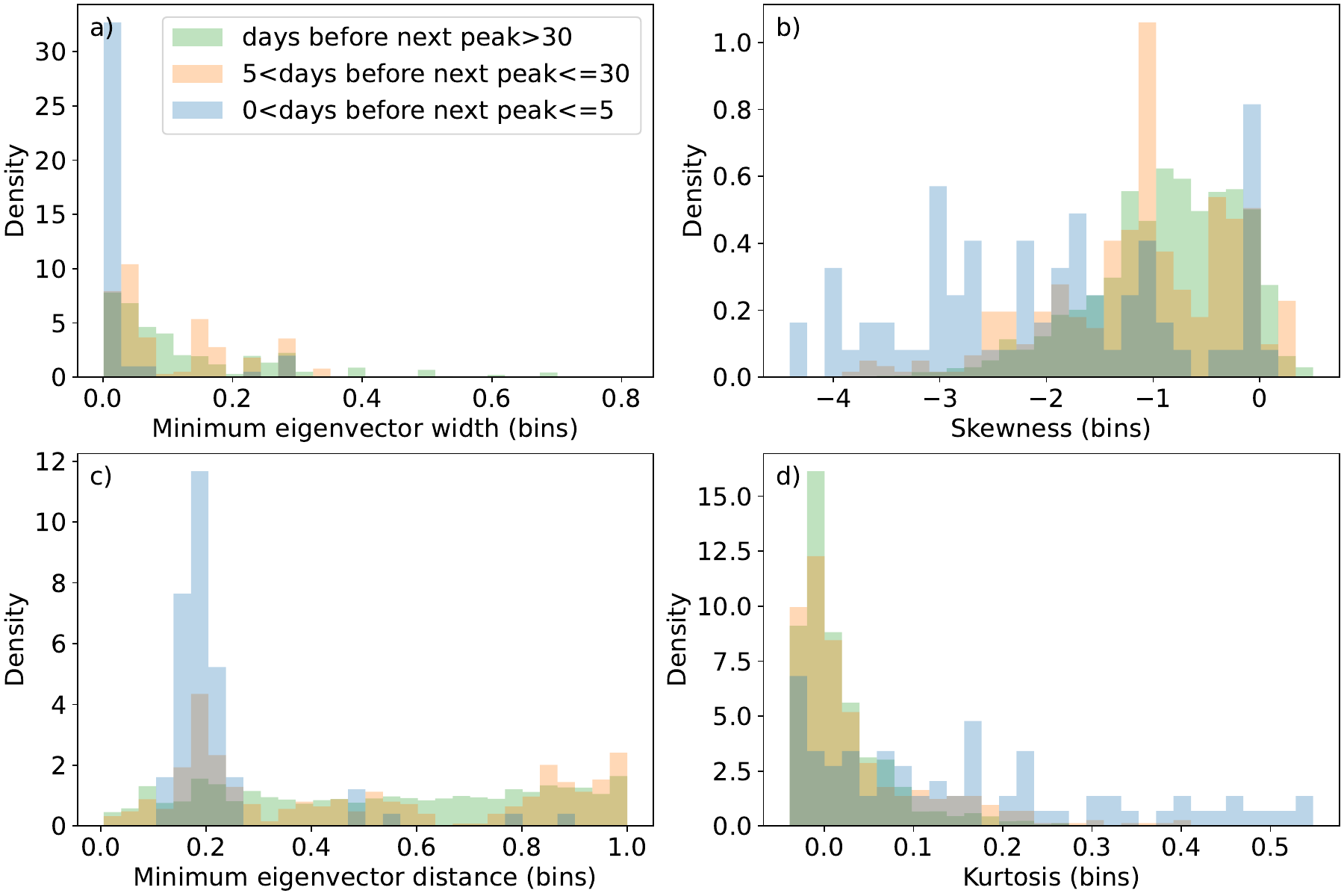}   
    \caption{a) Frequency density plot of minimum eigenvector widths at differing distances from the next extreme peak. In all panels, blue shows the distribution of values found between 0 and 5 days from a peak, orange between 5 and 30 days, and green for more than 30 days. b) Density plot of the skewness of the potential, c) density plot of the minimum eigenvector distance, and d) density plot of the kurtosis of the potential.}\label{fig:hist}
\end{figure}
If the width of the minimum eigenvectors of the system decrease in the time before an extreme event, we expect this to be an indicator of increased randomness, similar to what is seen for the eigenstates of random matrices when randomness is increased~\cite{mafi_transverse_2015}. In optics or hydrodynamics, this could also indicate more localised or stable states in the underlying system which have the potential to be enhanced by its nonlinear dynamics, leading to a large peak. However, it is unclear from this analysis if the analogy would extend this far. We observe a reduction in minimum eigenvector widths as an extreme peak is approached, suggesting that in this period the VIX is also displaying increased randomness. This shift in the statistical distribution of minimum eigenvector widths can be seen in figure \ref{fig:hist} a), with the distribution skewing much further towards 0 within five days before a peak. We also see a greater localisation of minimum eigenstates towards the peak as the days before the next peak decrease, displayed in panel c). Here, the distribution of eigenvector localisations appears to dramatically change from quite flat to narrow. Shifts in statistical distributions are also seen for the skewness (panel b)) and the kurtosis (panel d)) of the potential, suggesting a change in the nature of the VIX as a peak approaches. While these results support evidence of Anderson localisation in the eigenstates of VIX before a rogue wave, it is also clear that there are significant overlaps between the distributions covering different windows. It is important to note that in order to analyse the behaviour of the minimum eigenvector close to a rogue-wave-like peak, the time series must be processed up to and including the event, so it is identifiable. While these parameters could be analysed without knowledge of an oncoming extreme event, they are unlikely to be reliable predictors, as although the distributions shift as a peak approaches, they have significant overlap with the distributions far from a peak, too.

\begin{figure}[ht]
\centering
   \includegraphics[width=\linewidth]{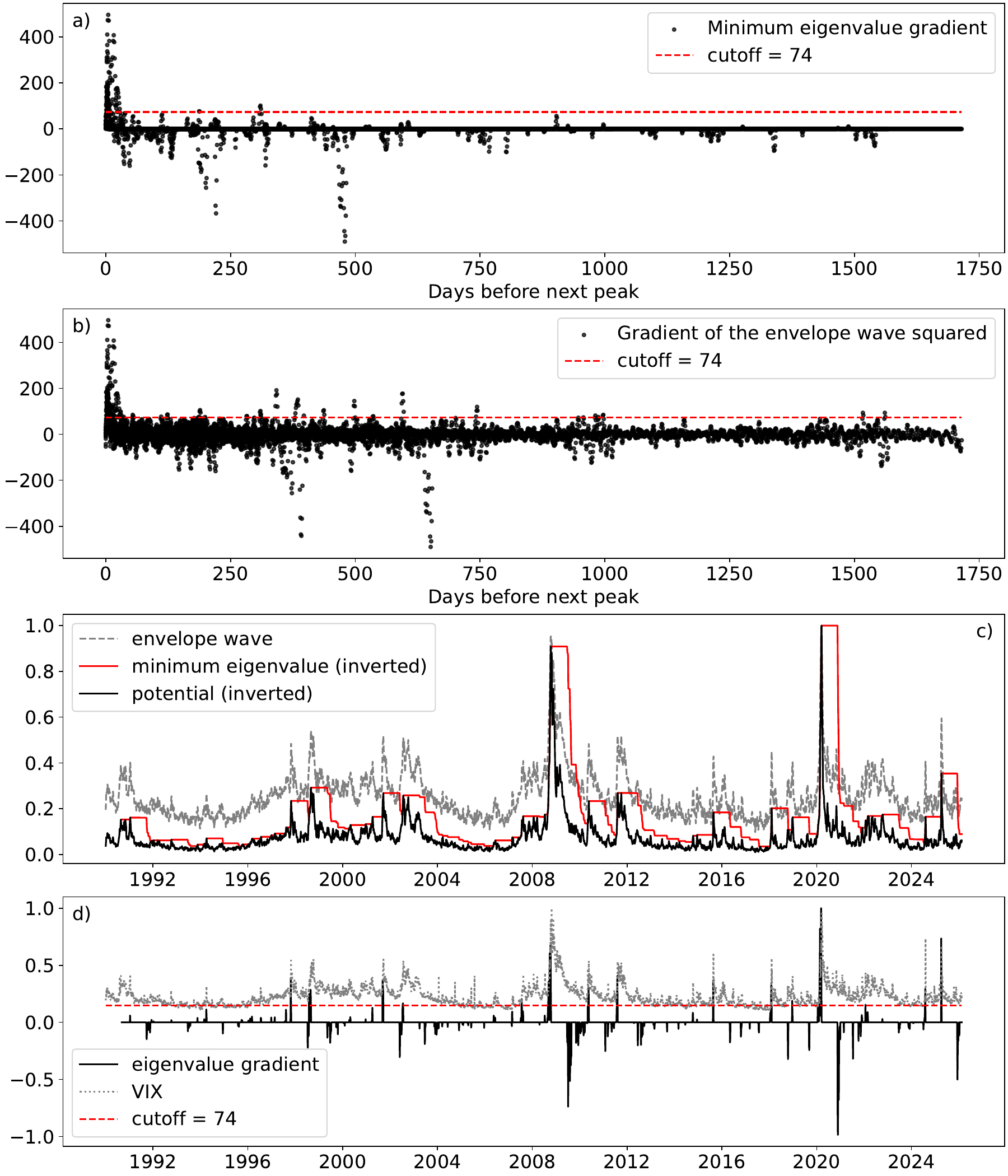}   
    \caption{a) Minimum eigenvalue gradient (black dots) for increasing number of days before the next extreme peak. The red dashed line marks the cut-off of 74, beyond which spikes are considered a reliable indicator of an oncoming peak. b) The gradient of the squared envelope wave (black dots) for an increasing number of days before the next peak. The same red line with the cut-off is shown. c) The envelope wave of the VIX (grey dashed), the potential (inverted, black), and the minimum eigenvalue (inverted, red). d) The VIX (grey dotted), the eigenvalue gradient, and the cut-off used to identify oncoming rogue waves.}\label{fig:gradients}
\end{figure}

However, when examining the envelope wave of the full series, a consistent signifier of an extreme event is the numerical gradient of the absolute value of the minimum eigenvalue of the potential, as defined in methods section. The minimum eigenvalue gradient is plotted against the days before the next identified rogue wave peak in the VIX in~\ref{fig:gradients} a). Comparing with the gradient of the envelope wave-squared in panel b), it is clear that the rate of change of the minimum eigenvalue provides a much less noisy distribution, with far smaller peaks far away from extreme events. This minimum eigenvalue will follow the minima of the square of the envelope wave within a given time window, see figure~\ref{fig:gradients} c). If the square of the envelope wave of the VIX grows very rapidly compared to its maximum value in the previous window of length L, it may indicate the onset of an extreme event. Notably, this will not work for the raw time series containing all noise and fluctuations - the long run, slowly-varying behaviour of the index must be extracted first. The squared-term also appears to be of importance, with similar results difficult to obtain for a weaker nonlinearity. The eigenvalue gradient (black) is plotted alongside the VIX (grey) in figure~\ref{fig:gradients} d). The dashed red line displays an arbitrarily identified cutoff of 74, beyond which the eigenvalue gradient consistently predicts an oncoming rogue wave peak. A caveat is that there exists significant information leakage in the system due to the shape of the envelope wave being highly dependent on the available data. In order to obtain a signal with predictive power, the full time series must not be processed in advance.

In the following subsections, we illustrate how our algorithm can be adapted to find signals in the VIX in simulated real time, as if new data appeared daily. Given that the full time series of the VIX has already been analysed, biasing knowledge of a suitable cutoff for the eigenvalue gradient, we conducted two out-of-sample tests: one on the VXO and one on VSTOXX. A historical subset of each dataset was used to calibrate the algorithm and choose appropriate parameter values based on observed performance, which were then tested on the remaining time series data.

\subsubsection*{Simulated real-time analysis}

As analysing the VIX as if receiving the data for each day in real time makes the algorithm more sensitive to small fluctuations, we increased the percentile threshold used to remove noise in Fourier space to 95 from 92.5 for all of the following results. It is important to stress that the wave envelope is only calculated up to the last available datapoint in the series, which prevents any information leakage. Our process for simulating the arrival of new data each day is described in the methods section. From the eigenvalue gradient signals received, we multiply the maximum eigenvalue gradient in the last ten days with the number of signals exceeding an arbitrary low threshold (50) to form a rogue wave warning indicator.

\begin{figure*}[ht]
\centering
   \includegraphics[width=0.9\linewidth]{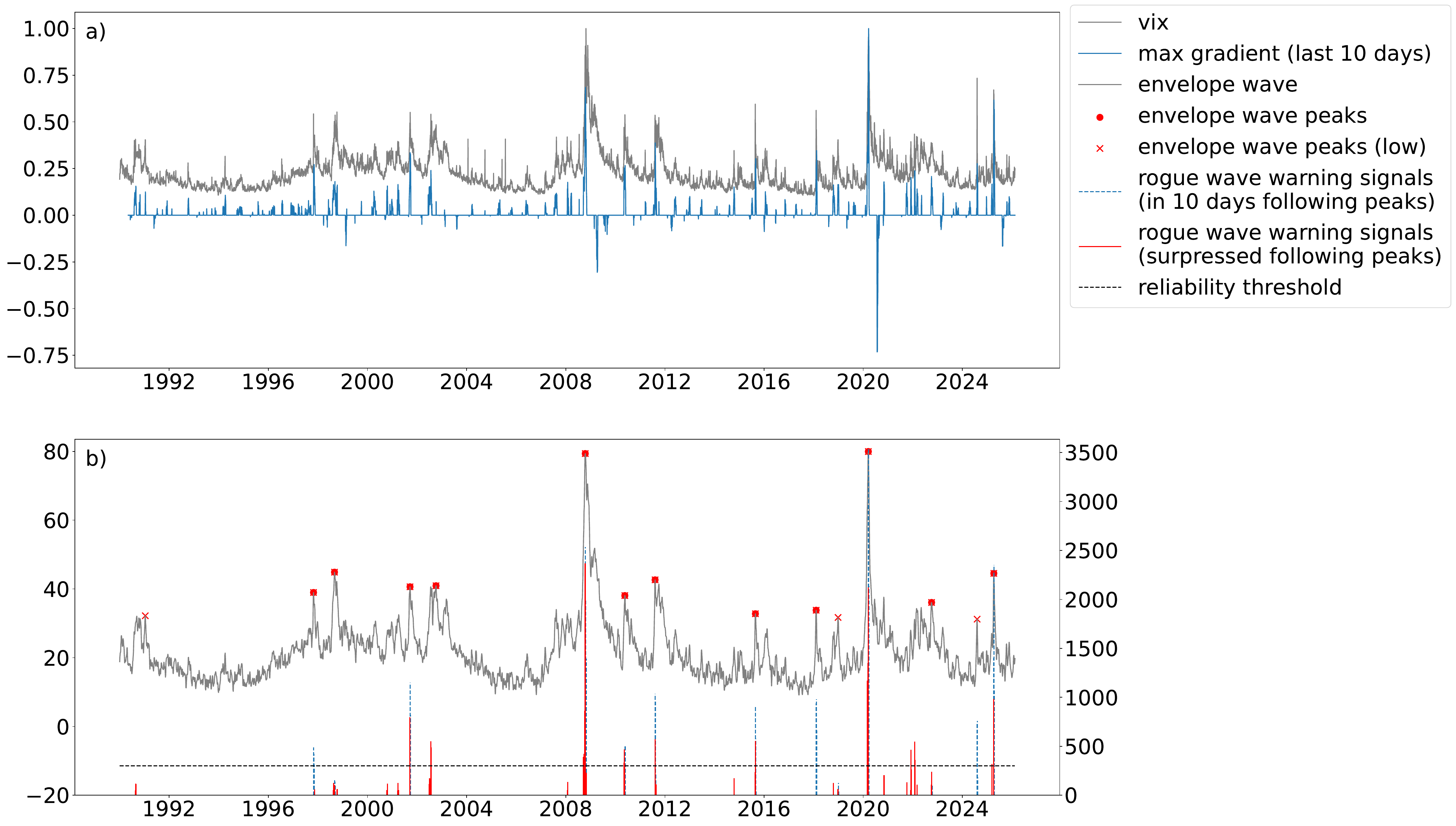}   
    \caption{a) The VIX (grey) and the maximum eigenvalue gradient found in the last ten days (blue). b) The envelope wave (grey), rogue wave peaks (red dots (2.5$\times$SWH) or crosses (2$\times$SWH)), and the rogue wave warning indicator (red vertical lines, except in the ten days following a peak where they are blue, dashed). }\label{fig:vixreal}
\end{figure*}

The results found when applying this methodology to the VIX can be seen in figure~\ref{fig:vixreal}. Panel a) displays the raw VIX data alongside the maximum eigenvalue gradient found in the ten days up to and including the point at which it is displayed. Panel b) displays the envelope wave for the full time series, used to assess whether the signals are reliable, with peaks exceeding 2.5 times the SWH marked by red dots, and peaks exceeding twice the SWH marked by red crosses. The red lines correspond to rogue wave warning signals, except in the 10 days following a peak, where the signals are instead marked by dashed blue lines -- this makes it clear when signals arrive after an extreme event. 

The methodology will perform poorly when limited data is available early on in the time series, due to the inability of the signal processing methods to correctly identify a suitable wave envelope. However, once a good fit is achieved the rogue wave warning signals perform extremely well, only failing to identify the 2018 VIX peak in advance, the rise of which happens over an extremely short timescale relative to the granularity of data points (over approximately two days for daily data, see supplementary figure S3). Signal counts and signal strengths are also shown, with the March 2020 peak shown for comparison. The earliest rogue wave warning in 2008 corresponds to the collapse of the Lehman Brothers on Monday the 15th of September 2008, and is identified by the algorithm a few days after on Thursday the 18th. The algorithm therefore cannot predict financial crashes, but identifies events in the VIX which are likely to mark the beginning of a rogue wave once it is able to resolve them. Nevertheless, this warning on the 18th appears a full 18 trading days total before the peak of the VIX envelope on the 14th of October (the date of the envelope peak changes compared to the peak identified in supplementary figure S1 due to increased removal of noise). We examine the strength of the rogue wave warning indicators in relation to each major peak, identifying a threshold of 300 through repeated trials as a clear sign that a rogue-like peak is approaching while minimising false alerts. We then use this threshold to determine when a rogue wave warning signal is considered reliable based on the available results. The performance statistics based on this threshold for the first signals which appear can be seen in table~\ref{tab:vix}. Detailed data on the appearance of signals and locations of peaks can be found in supplementary table S1.

\begin{table}[ht]
\centering
\begin{tabular}{|l|l|l|}
\hline
Condition & \%  \tiny{($>$ 2.5 SWH)}& \% \tiny{($>$ 2 SWH)} \\
\hline
Peaks detected & 75.0 & 60.0 \\
\hline
False signals & 0  & 0 \\
\hline
Precision & 84.2  & 84.2 \\
\hline
Precision (exc. late signals) & 94.1 & 100  \\
\hline
\end{tabular}
\caption{\label{tab:vix} Peak detection statistics for the VIX. Precision is the percentage of signals which can be considered a reliable indicator of an oncoming extreme event. Details on how statistics were calculated can be found in the methods section.}
\end{table}

\subsection*{Out-of-sample test}

In order to test whether the success of our methods extends beyond the VIX, we perform two out-of-sample tests, one for the VXO and one for VSTOXX, by calculating the rogue wave warning indicators for a historical subset of the data. We adjust the parameters of the system and examine the strength of the rogue wave warning indicator, repeating the in-sample tests until a threshold over which the indicator reliably corresponds to an oncoming extreme event is identified. We then test whether indicators found for the out-of-sample subset of the data pass this threshold. These thresholds will be dependent on the system being investigated. The moving window size remains fixed at $L=80$ trading days (other values were considered but 80 trading days performed best in all cases), with the low signal cut-off of 50 kept when calculating the warning indicator (this low number relative to the numerical eigenvalue gradients recorded is chosen simply to eliminate small fluctuations). 

\begin{figure}[ht]
\centering
   \includegraphics[width=\linewidth]{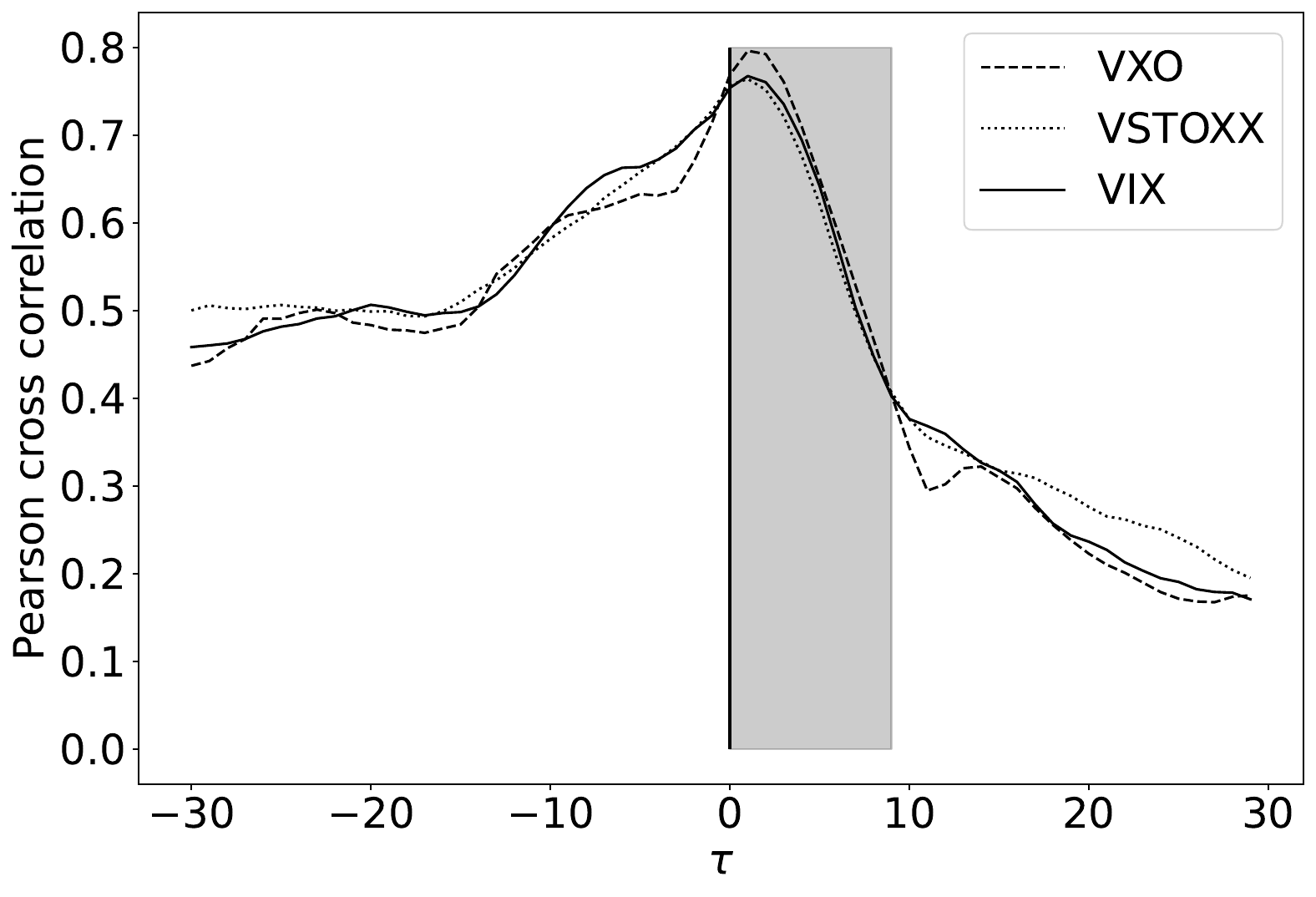}   
    \caption{Pearson cross correlation between the maximum eigenvalue gradient for each volatility index and the height of the volatility index.}\label{fig:crosscorr}
\end{figure}

For all three volatility indices, the maximum value of numerical eigenvalue gradient (for values > 0.0001) remained a leading indicator of the index, as seen in figure \ref{fig:crosscorr}. The peak of the cross correlation appearing after $\tau=0$ is expected given that the maximum eigenvalue gradient is taken for the ten days up to and including the date recorded; this also explains why the decline in correlation after 0 is less severe after approximately ten days in all cases. 

\subsubsection*{VXO}

\begin{figure*}[ht]
\centering
   \includegraphics[width=\linewidth]{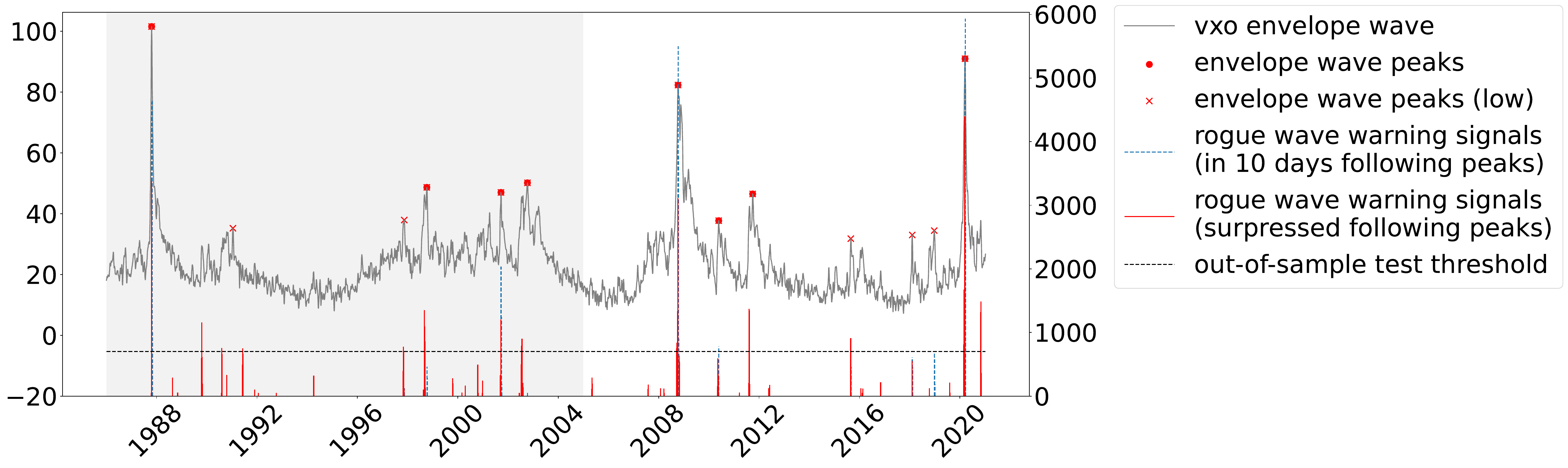}   
    \caption{ The envelope wave of the VXO index (grey line), the in-sample area (grey shading), rogue wave peaks (red dots (2.5$\times$SWH) or crosses ($>$2$\times$SWH), the rogue wave warning indicator (red vertical lines, except for the ten days following an extreme peak where it is displayed using blue dashed lines), and the out of sample test threshold (grey dashed). }\label{fig:vxoreal}
\end{figure*}

The results for the VXO index can be seen in figure \ref{fig:vxoreal}, in the same layout as figure \ref{fig:vixreal} b) with the addition of a shaded area covering the in-sample period, and a dashed grey line in panel b) indicating the out-of-sample test threshold over which rogue wave warning signals are considered reliable. The threshold identified after repeated trials in the in-sample period is equal to 700 for the VXO; this number was chosen to maximise inclusion of reliable signals while minimising noise. Only one major peak is missed by this threshold in the out-of-sample test, but is arguably still marked by a significant spike in the rogue wave warning indicator. There are several false alerts early in the series, but this is again due to the limited data available when resolving the envelope wave, and precision improves dramatically after a few years. The in-sample cut-off point was the last trading day in 2004. The out-of-sample period spans the beginning of 2005 to the end of 2020.

A significantly high rogue wave warning spike is seen just after the major peak in March 2020, which we show an expanded view of in supplementary figure S4. While not a ``rogue wave'' in its own right, it corresponds to a sharp rise in the VXO. The maximum eigenvalue gradient is plotted here for the day of the event vs the day the signal is received, in order for the characteristics of larger and smaller events to be compared.

Table \ref{tab:vxo} displays the overall performance of the rogue wave warning indicator for the VXO, for the full time series. More detailed data on the appearance of signals can be found in supplementary table S2. False signals refer to indicators which pass the threshold, but which appear more than three windows ($L=80$) before the next identified peak: this is a large amount of time, but the most extreme events can start several weeks in advance. The precision refers to the percentage of signals which correspond to an extreme event. It is clear that the 2010 signals fall slightly short of the threshold, the height of which was influenced by the false alerts seen very early on in the VXO time series. The only false signal after 1996 is that seen in supplementary figure S4.

\begin{table}[ht]
\centering
\begin{tabular}{|l|l|l|}
\hline
Condition & \%  \tiny{($>$ 2.5 SWH)}& \% \tiny{($>$ 2 SWH)} \\
\hline
Peaks detected & 87.5 & 76.9 \\
\hline
False signals & 15.8  & 15.8 \\
\hline
Precision & 63.2  & 73.7  \\
\hline
Precision (exc. late signals) & 66.7 & 82.4  \\
\hline
\end{tabular}
\caption{\label{tab:vxo} Peak detection statistics for the VXO. Precision is the percentage of signals which can be considered a reliable indicator of an oncoming extreme event. Details on how statistics were calculated can be found in the methods section.}
\end{table}

\subsubsection*{VSTOXX}

\begin{figure*}[ht]
\centering
   \includegraphics[width=\linewidth]{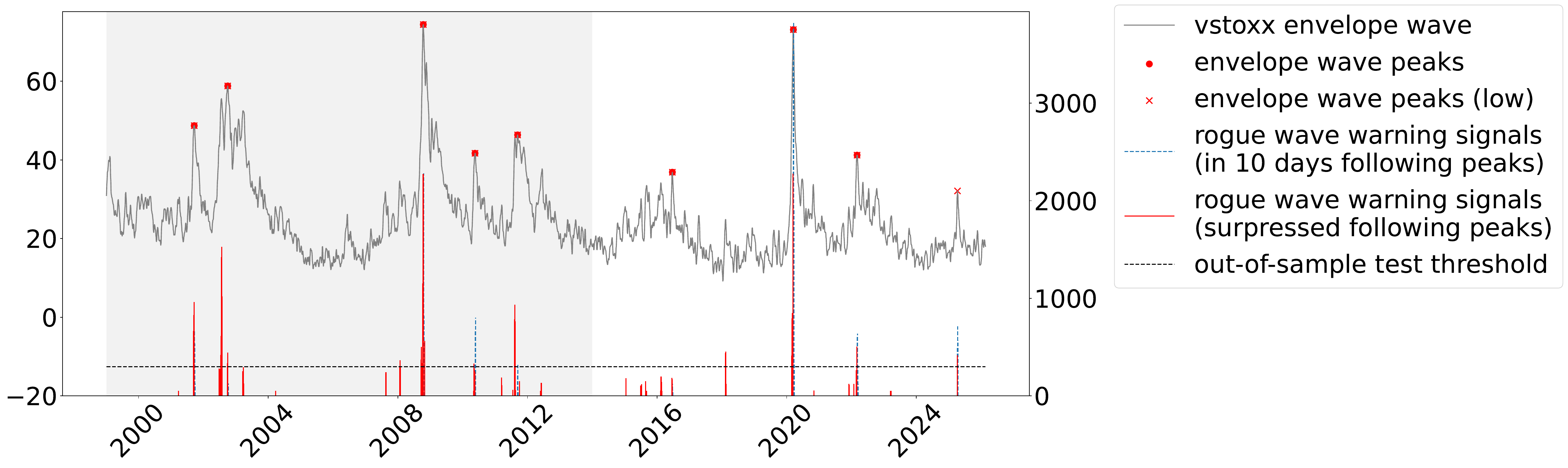}   
    \caption{ The envelope wave of the VSTOXX index (grey line), the in-sample area (grey shading), rogue wave peaks (red dots (2.5$\times$SWH) or crosses ($>$2$\times$SWH), the rogue wave warning indicator (red vertical lines, except for the ten days following an extreme peak where it is displayed using blue dashed lines), and the out of sample test threshold (grey dashed). }\label{fig:vstoxxreal}
\end{figure*}

The out-of-sample test results for the VSTOXX index are displayed in figure \ref{fig:vstoxxreal}. Here, the threshold chosen based on the results of repeated trials within the shaded in-sample period is equal to 300. The results for this index are very precise ($>$70\%), with the 2016 peak missed, but given its more rounded shape and less rapid rise and fall it is not surprising that less extreme indicators are seen. 

The performance statistics for the full time series can be seen in table \ref{tab:vstoxx}. There is a false signal in 2008 which appears more than 10 days following the peak, but is arguably unsurprising given the long length of the crisis, and another in 2018 which corresponds to the recovery following a sharp dip in the index. More detailed data on the first rogue warning signals which appear in a series of repeated signals can be found in supplementary table S3.

\begin{table}[ht]
\centering
\begin{tabular}{|l|l|l|}
\hline
Condition & \% \tiny{($>$ 2.5 SWH)}& \% \tiny{($>$ 2 SWH)}  \\
\hline
Peaks detected & 87.5 & 88.9 \\
\hline
False signals & 11.8  &  11.8 \\
\hline
Precision & 70.6 &  76.5 \\
\hline
Precision (exc. late signals) & 80.0 & 86.7  \\
\hline
\end{tabular}
\caption{\label{tab:vstoxx} Peak detection statistics for the VSTOXX. Precision is the percentage of signals which can be considered a reliable indicator of an oncoming extreme event. Details on how statistics were calculated can be found in the methods section.}
\end{table}

\section*{Discussion}

The practical interpretation of why the rate of change of the system's lowest eigenvalue allows us to identify the onset of an extreme event is as follows: the minimum eigenvalue closely tracks the minima of the potential (the squared envelope of the volatility index) within a given time window. This can be seen in its square-like shape in figure \ref{fig:gradients} c). A sudden rise in the slowly-varying behaviour of the volatility index compared to the previous 80 trading days (or whatever the chosen window length L is set to), captured by the envelope, will therefore correspond to a sharp increase in this eigenvalue gradient. 

We therefore believe what we are observing corresponds to a regime change in these systems, where the slowly-varying part of the system begins to take on nonlinear behaviour leading to a cascade. This could be due to the underlying stock option prices which make up each index beginning to depend more on their own recent history and also on each other. However, further exploration of our method applied to both composite indices and individual stock prices is needed to understand how it relates to the underlying system dynamics.

It is difficult to compare our findings and their success in identifying the onset of extreme events to current models in finance, due to differing approaches to modelling and estimating success. The maximum eigenvalue gradient as a leading indicator (negative values excluded) is fairly highly correlated with the values of implied volatility 10 days ahead, providing some information on the likely magnitude and direction of volatility in the future. Models which aim to forecast volatility can be very successful: 10 day ahead models are able to predict approximately 65-88\% of directional changes in implied volatility~\cite{degiannakis_forecasting_2018}, but it is hard to say whether this translates to the ability to identify a true oncoming extreme event. Precursory patterns before an extreme financial event were previously identified by Sornette, Johansen, and Bouchaud in 1996~\cite{sornette_stock_1996}, leading to much research into Log-Periodic Power Law Models as a forecasting tool. However, a recent study which integrated the use of artificial intelligence into these models achieved a maximum precision of 50.88\% when identifying crashes~\cite{lee_more_2025}. Contrasting with the rogue wave warning indicator, at most times it will have no value at all, and will spike before all peaks -- where it misses, it is due to the failure of the process to separate short-term fluctuations from long term behaviour, or due to the granularity of the data meaning a peak could not be identified in advance. Strong spikes in the out-of-sample periods do not always correspond to identified rogue waves, but do always correspond to sudden rises in volatility, with repeated or growing signals clearly marking a true escalation. Some trade-off between precision and peak detection performance is clear, with a certain amount of false alerts or signals for lower peaks perhaps a reasonable cost for ensuring all extreme events are detected. This work is also limited by the current estimation of the parameters such as the window size, signal cut-off, and threshold, which are obtained by repeated trials when studying the in-sample portions of datasets. As the strength of the eigenvalue gradient before an extreme event varies significantly between systems, further analysis is needed to see if a general rule for threshold identification can be found, eliminating arbitrary choices of parameters.

We believe these initial results are extremely promising given the novelty of this method, and with further research could provide an even more reliable and powerful tool for identifying extremes in volatility indices and similar financial time series. The future performance of our method could be influenced by the actions of traders if put into use, as an indication that an extreme in financial volatility is approaching could increase or decrease future volatility, depending on behaviour.

\section*{Methods}

\subsection*{Signal processing}

To isolate long-run changes in the volatility indices, we first remove noise by using Fast Fourier Transform to identify and exclude the least common frequencies which appear in the signal. We then use a Hilbert transform to isolate the analytic signal, taking the absolute value to extract the envelope wave. These methods make assumptions of the periodicity of the signal which can lead to misleading spikes at the end of a series after processing. As we want to potentially pick up genuine sharp signals at the end of each series during the simulated real-time analysis, we mirror the signal; this ensures the start and end points of the signal are not modified to account for the assumption of periodicity and provide a more realistic wave envelope at the end points. We find this method works well to prevent the introduction of errors during the simulated real-time analysis.

We construct a noise filter by removing any peaks which fall below either the 92.5th percentile or the 95th percentile of peaks in Fourier space. The amplitude envelope of the wave is extracted by taking the absolute value of the analytic signal, obtained using the Hilbert transform. 

\subsection*{Rogue wave identification}

In order to identify wave heights in the volatility series, we first identify all peaks in the signal using the find\_peaks SciPy function, and extract the prominence of these peaks. The prominence is the vertical distance from a peak to its lowest contour line, and therefore we consider it a reasonable analogy to the definition of wave height (peak to trough distance) used in ocean science when considering only the envelope wave of the series. We identify the significant wave height of each index as the average of the top third of its wave prominences. We then identify peaks greater than $2\times$ and $2.5\times$ the SWH, which can be seen in figure \ref{fig:peaks} for the VIX. Peaks found after signal processing in order to extract the envelope wave of the signal are shown in red (x's for twice the SWH, dots for 2.5), and are far less frequent than those found using the raw dataset, shown with grey x's for peaks exceeding $2.5\times$ SWH. 

The full time series of each volatility index is used to identify rogue-like peaks, separate from the simulated real-time analysis. It is then used to check if the results of the rogue wave warnings found in the real-time analysis align with the identified extreme events. However, the available data will have a significant effect on which peaks are identified as rogue waves, with the distribution changing whenever each series is extended. The significant wave height could be identified using smaller windows of data, however the history of extreme financial events is pivotal in assessing the significance of an extreme event in the VIX VXO, or VSTOXX, so the authors feel it is better to assess the entire time series when estimating rogue waves.

\subsection*{Potential and eigenvalues}

The Nonlinear Schr\"odinger equation for a classical wave field can be written
\begin{equation}\label{eq:sch}
i\frac{\partial\psi}{\partial z} = -\frac{1}{2}\frac{\partial^2\psi}{\partial t^2} + \kappa |\psi|^2\psi 
\end{equation}
where $\kappa$ is a constant and $\psi$ is a complex field. We can look for evidence of localised states in this system by examining the eigenstates and eigenvalues of the linear equation:
\begin{equation}\label{eq:lin}
\lambda \psi = -\frac{1}{2}\frac{\partial^2\psi}{\partial t^2} + U\psi,
\end{equation}
where $U$ follows the shape of $- |\psi|^2$. We make this choice in analogy with the linear equation found in optics when spatio-temporal modulation causes the refractive index of a system to follow the shape of the Kerr nonlinearity~\cite{saleh_anderson_2017}. After normalising the envelope amplitude and measuring time in units of one trading day, we define the dimensionless potential as:
\begin{equation}\label{eq:potential}
U = - |\psi|^2.
\end{equation}
As $\psi$ is a discrete time series of points within a given time window of length L, by using the finite difference approximation for the second derivative,
\begin{equation}
\frac{\partial^2\psi}{\partial t^2} = \frac{\psi_{t+1} - 2\psi_t +\psi_{t-1}}{\Delta t^2} ,
\end{equation}
the instantaneous eigenvalues and eigenvectors can easily be found by combining with the potential to form a matrix operator, and solving for the eigensystem.

\subsection*{Localisation}

The eigenvector width and distance seen in figure~\ref{fig:hist} are based on the average width of the peak of the absolute minimum eigenvector using the second-moment method~\cite{mafi_transverse_2015}, and the distance between its maximum absolute point and the end of the window of data used for its calculation. 

The width is given by the simple formula:

\begin{equation}
w(v) = \sqrt \frac{\int (|v(t)|^2(t-t_0)^2)}{\int (|v(t)|^2)},
\end{equation}

where the trapezoid rule is used to calculate the definite integral, and $t_0$ corresponds to the location of the maximum argument of $|v|$. 

\subsection*{Simulating real-time analysis}

Our process for simulating the real-time analysis of financial volatility data is as follows:

\begin{itemize}
\item A starting length of data equal to the window length L plus 9 additional days is needed to run the analysis. For each new data point in the series passed to the algorithm, the envelope wave is extracted based on the data up to and including that point.
\item For the ten data points (each data point corresponding to a single trading day) $j \in [i-9,i]$ up to and including the last point in this envelope wave, $i$, the eigenvalues and eigenvectors of equation~\ref{eq:lin} are calculated for the window $j-L:j$. 
\item The numerical eigenvalue gradient is calculated based on the absolute values of the lowest eigenvalues found for each of these ten days. 
\item The maximum eigenvalue gradient found within these ten days and the corresponding date are extracted and stored, as is the count of the number of days where the eigenvalue gradient exceeds an arbitrary minimum threshold. A minimum threshold of 50 was used for all examples presented here.
\item The rogue wave warning signals are calculated by multiplying the number of signals which exceed the threshold by the maximum eigenvalue gradient.
\item The rogue wave warning date is stored as the last day of data passed to the algorithm,  which we refer to as the ``signal date''. However, this date can be compared to the date of the maximum eigenvalue gradient, which we refer to as the ``event date''.
\end{itemize}

\subsection*{Performance statistics}

The statistics seen in tables~\ref{tab:vix}, \ref{tab:vxo}, and \ref{tab:vstoxx} are based on the first rogue wave warning signals received in a continuous stream of signals. A peak is considered detected if a first signal appears in the preceding three windows before a peak and it is more than one window away from the last peak. This is a generous amount of time, however the widths of the peaks can be extremely large. Missed peaks are labelled ``no signal''. The percentage of peaks detected is therefore simply the number of peaks minus the number of no-signals, divided by the number of peaks.

Due to the variation of peak widths, signals are considered false alerts or ``false signals'' if they appear more than three windows before the next peak. The percentage of false signals is given by the number of false signals divided by the total number of signals.

Signals are considered a late signal if they appear within the ten days following a peak, due to the fact the previous ten days are used when calculating rogue wave warning signals. Signals are considered a low signal if the peak they appear before is more than twice the SWH but not more than 2.5 times the SWH. 

Precision is calculated by considering how many of these first signals can be considered useful; it is the percentage of signals which are not false, low, or late, when considering only major peaks:
\begin{equation}
\mathrm{Precision} = \frac{\mathrm{True \,Positives}}{\mathrm{True \, Positives} + \mathrm{False \, Positives}}.
\end{equation}
 We also calculate the percentage ignoring late signals, as it is quite easy to confirm whether the event they correspond to has passed or not.

These statistics were compiled using very simple rules, and in reality determining whether a rogue wave warning signal is valid or not depends on the specific conditions of the event in question: when does it start and end, how long does it last, how rapidly does the value of the envelope wave escalate?

\section{Author contributions}

FB formulated theory, and designed and coordinated initial research. FB and OL conducted preliminary analysis on the VIX identifying evidence of Anderson localisation and the importance of the eigenvalue gradient. RH designed and performed follow up research including peak identification, extreme event statistics, real-time analysis, and out-of-sample testing of the VXO and VSTOXX indices with support from FB and OL. RH wrote the article with input from FB and OL.

\section{Acknowledgements}

RH would like to thank P. Klimek for discussions on information leakage and leading indicators, and for reviewing our manuscript; J.P. Bouchaud for discussions on stock price correlations and fractal nature of volatility; and A. Bugaenko  and L. Ialongo for sharing relevant literature. On behalf of the Supply Chain Intelligence Institute Austria (ASCII), RH acknowledges financial support from the Austrian Federal Ministry for Economy, Energy and Tourism (BMWET) and the Federal State of Upper Austria.

\bibliography{references}

\begin{thebibliography}{10}
\urlstyle{rm}
\expandafter\ifx\csname url\endcsname\relax
  \def\url#1{\texttt{#1}}\fi
\expandafter\ifx\csname urlprefix\endcsname\relax\def\urlprefix{URL }\fi
\expandafter\ifx\csname doiprefix\endcsname\relax\def\doiprefix{DOI: }\fi
\providecommand{\bibinfo}[2]{#2}
\providecommand{\eprint}[2][]{\url{#2}}

\bibitem{guerrero_multiplicative_2020}
\bibinfo{author}{Guerrero, F.~G.} \& \bibinfo{author}{Garcia-Baños, A.}
\newblock \bibinfo{journal}{\bibinfo{title}{Multiplicative processes as a
  source of fat-tail distributions}}.
\newblock {\emph{\JournalTitle{Heliyon}}} \textbf{\bibinfo{volume}{6}},
  \bibinfo{pages}{e04266}, \doiprefix\url{10.1016/j.heliyon.2020.e04266}
  (\bibinfo{year}{2020}).

\bibitem{sornette_stock_1996}
\bibinfo{author}{Sornette, D.}, \bibinfo{author}{Johansen, A.} \&
  \bibinfo{author}{Bouchaud, J.-P.}
\newblock \bibinfo{journal}{\bibinfo{title}{Stock {Market} {Crashes},
  {Precursors} and {Replicas}}}.
\newblock {\emph{\JournalTitle{J. Phys. I France}}}
  \textbf{\bibinfo{volume}{6}}, \bibinfo{pages}{167--175},
  \doiprefix\url{10.1051/jp1:1996135} (\bibinfo{year}{1996}).

\bibitem{lee_more_2025}
\bibinfo{author}{Lee, G.}, \bibinfo{author}{Jeong, M.}, \bibinfo{author}{Park,
  T.} \& \bibinfo{author}{Ahn, K.}
\newblock \bibinfo{journal}{\bibinfo{title}{More than ex-post fitting:
  log-periodic power law and its {AI}-based classification}}.
\newblock {\emph{\JournalTitle{Humanit Soc Sci Commun}}}
  \textbf{\bibinfo{volume}{12}}, \bibinfo{pages}{1664},
  \doiprefix\url{10.1057/s41599-025-05920-7} (\bibinfo{year}{2025}).

\bibitem{dudley_rogue_2019}
\bibinfo{author}{Dudley, J.~M.}, \bibinfo{author}{Genty, G.},
  \bibinfo{author}{Mussot, A.}, \bibinfo{author}{Chabchoub, A.} \&
  \bibinfo{author}{Dias, F.}
\newblock \bibinfo{journal}{\bibinfo{title}{Rogue waves and analogies in optics
  and oceanography}}.
\newblock {\emph{\JournalTitle{Nat Rev Phys}}} \textbf{\bibinfo{volume}{1}},
  \bibinfo{pages}{675--689}, \doiprefix\url{10.1038/s42254-019-0100-0}
  (\bibinfo{year}{2019}).

\bibitem{haver_possible_2004}
\bibinfo{author}{Haver, S.}
\newblock \bibinfo{journal}{\bibinfo{title}{A {Possible} {Freak} {Wave} {Event}
  {Measured} at the {Draupner} {Jacket} {January} 1 1995}}.
\newblock {\emph{\JournalTitle{Actes de colloques-IFREMER}}}
  \textbf{\bibinfo{volume}{39}} (\bibinfo{year}{2004}).

\bibitem{solli_optical_2008}
\bibinfo{author}{Solli, D.}, \bibinfo{author}{Ropers, C.},
  \bibinfo{author}{Koonath, P.} \& \bibinfo{author}{Jalali, B.}
\newblock \bibinfo{journal}{\bibinfo{title}{Optical rogue waves}}.
\newblock {\emph{\JournalTitle{Nature}}} \textbf{\bibinfo{volume}{450}},
  \bibinfo{pages}{1054--7}, \doiprefix\url{10.1038/nature06402}
  (\bibinfo{year}{2008}).

\bibitem{gutenberg_frequency_1944}
\bibinfo{author}{Gutenberg, B.} \& \bibinfo{author}{Richter, C.~F.}
\newblock \bibinfo{journal}{\bibinfo{title}{Frequency of earthquakes in
  {California}}}.
\newblock {\emph{\JournalTitle{Bulletin of the Seismological Society of
  America}}} \textbf{\bibinfo{volume}{34}}, \bibinfo{pages}{185--88},
  \doiprefix\url{10.1785/BSSA0340040185} (\bibinfo{year}{1944}).

\bibitem{hutt_synergetics_2020}
\bibinfo{editor}{Hutt, A.} \& \bibinfo{editor}{Haken, H.} (eds.)
  \emph{\bibinfo{title}{Synergetics}} (\bibinfo{publisher}{Springer US},
  \bibinfo{address}{New York, NY}, \bibinfo{year}{2020}).

\bibitem{liu_statistical_1999}
\bibinfo{author}{Liu, Y.} \emph{et~al.}
\newblock \bibinfo{journal}{\bibinfo{title}{The statistical properties of the
  volatility of price fluctuations}}.
\newblock {\emph{\JournalTitle{Phys. Rev. E}}} \textbf{\bibinfo{volume}{60}},
  \bibinfo{pages}{1390--1400}, \doiprefix\url{10.1103/PhysRevE.60.1390}
  (\bibinfo{year}{1999}).
\newblock \bibinfo{note}{ArXiv:cond-mat/9903369}.

\bibitem{parsons_global_2002}
\bibinfo{author}{Parsons, T.}
\newblock \bibinfo{journal}{\bibinfo{title}{Global {Omori} law decay of
  triggered earthquakes: {Large} aftershocks outside the classical aftershock
  zone}}.
\newblock {\emph{\JournalTitle{J. Geophys. Res.}}}
  \textbf{\bibinfo{volume}{107}}, \doiprefix\url{10.1029/2001JB000646}
  (\bibinfo{year}{2002}).

\bibitem{toth_studies_2009}
\bibinfo{author}{Toth, B.}, \bibinfo{author}{Kertesz, J.} \&
  \bibinfo{author}{Farmer, J.~D.}
\newblock \bibinfo{journal}{\bibinfo{title}{Studies of the limit order book
  around large price changes}}.
\newblock {\emph{\JournalTitle{Eur. Phys. J. B}}}
  \textbf{\bibinfo{volume}{71}}, \bibinfo{pages}{499--510},
  \doiprefix\url{10.1140/epjb/e2009-00297-9} (\bibinfo{year}{2009}).
\newblock \bibinfo{note}{ArXiv:0901.0495 [q-fin]}.

\bibitem{lillo_power-law_2003}
\bibinfo{author}{Lillo, F.} \& \bibinfo{author}{Mantegna, R.~N.}
\newblock \bibinfo{journal}{\bibinfo{title}{Power-law relaxation in a complex
  system: {Omori} law after a financial market crash}}.
\newblock {\emph{\JournalTitle{Phys. Rev. E}}} \textbf{\bibinfo{volume}{68}},
  \bibinfo{pages}{016119}, \doiprefix\url{10.1103/PhysRevE.68.016119}
  (\bibinfo{year}{2003}).

\bibitem{pasquini_multiscale_1999}
\bibinfo{author}{Pasquini, M.} \& \bibinfo{author}{Serva, M.}
\newblock \bibinfo{journal}{\bibinfo{title}{Multiscale behaviour of volatility
  autocorrelations in a financial market}}.
\newblock {\emph{\JournalTitle{Economics Letters}}}
  \textbf{\bibinfo{volume}{65}}, \bibinfo{pages}{275--279},
  \doiprefix\url{10.1016/S0165-1765(99)00159-7} (\bibinfo{year}{1999}).

\bibitem{ghashghaie_turbulent_1996}
\bibinfo{author}{Ghashghaie, S.}, \bibinfo{author}{Breymann, W.},
  \bibinfo{author}{Peinke, J.}, \bibinfo{author}{Talkner, P.} \&
  \bibinfo{author}{Dodge, Y.}
\newblock \bibinfo{journal}{\bibinfo{title}{Turbulent cascades in foreign
  exchange markets}}.
\newblock {\emph{\JournalTitle{Nature}}} \textbf{\bibinfo{volume}{381}},
  \bibinfo{pages}{767--770}, \doiprefix\url{10.1038/381767a0}
  (\bibinfo{year}{1996}).

\bibitem{ivancevic_adaptive-wave_2010}
\bibinfo{author}{Ivancevic, V.~G.}
\newblock \bibinfo{journal}{\bibinfo{title}{Adaptive-{Wave} {Alternative} for
  the {Black}-{Scholes} {Option} {Pricing} {Model}}}.
\newblock {\emph{\JournalTitle{Cognitive Computation}}}
  \textbf{\bibinfo{volume}{2}}, \bibinfo{pages}{17--30},
  \doiprefix\url{10.1007/s12559-009-9031-x} (\bibinfo{year}{2010}).

\bibitem{yan_financial_2010}
\bibinfo{author}{Yan, Z.}
\newblock \bibinfo{journal}{\bibinfo{title}{Financial {Rogue} {Waves}}}.
\newblock {\emph{\JournalTitle{Commun. Theor. Phys.}}}
  \textbf{\bibinfo{volume}{54}}, \bibinfo{pages}{947},
  \doiprefix\url{10.1088/0253-6102/54/5/31} (\bibinfo{year}{2010}).

\bibitem{aubrun_identifying_2025}
\bibinfo{author}{Aubrun, C.}, \bibinfo{author}{Morel, R.},
  \bibinfo{author}{Benzaquen, M.} \& \bibinfo{author}{Bouchaud, J.-P.}
\newblock \bibinfo{journal}{\bibinfo{title}{Identifying new classes of
  financial price jumps with wavelets}}.
\newblock {\emph{\JournalTitle{Proceedings of the National Academy of
  Sciences}}} \textbf{\bibinfo{volume}{122}}, \bibinfo{pages}{e2409156121},
  \doiprefix\url{10.1073/pnas.2409156121} (\bibinfo{year}{2025}).

\bibitem{zakharov_stability_1972}
\bibinfo{author}{Zakharov, V.~E.}
\newblock \bibinfo{journal}{\bibinfo{title}{Stability of periodic waves of
  finite amplitude on the surface of a deep fluid}}.
\newblock {\emph{\JournalTitle{J Appl Mech Tech Phys}}}
  \textbf{\bibinfo{volume}{9}}, \bibinfo{pages}{190--194},
  \doiprefix\url{10.1007/BF00913182} (\bibinfo{year}{1972}).

\bibitem{chiao_self-trapping_1964}
\bibinfo{author}{Chiao, R.~Y.}, \bibinfo{author}{Garmire, E.} \&
  \bibinfo{author}{Townes, C.~H.}
\newblock \bibinfo{journal}{\bibinfo{title}{Self-{Trapping} of {Optical}
  {Beams}}}.
\newblock {\emph{\JournalTitle{Phys. Rev. Lett.}}}
  \textbf{\bibinfo{volume}{13}}, \bibinfo{pages}{479--482},
  \doiprefix\url{10.1103/PhysRevLett.13.479} (\bibinfo{year}{1964}).

\bibitem{kato_nonlinear_2005}
\bibinfo{author}{Kato, T.}
\newblock \bibinfo{title}{Nonlinear {Schrödinger} equations}.
\newblock In \bibinfo{editor}{Holden, H.} \& \bibinfo{editor}{Jensen, A.}
  (eds.) \emph{\bibinfo{booktitle}{Schrödinger {Operators}}},
  \bibinfo{pages}{218--263} (\bibinfo{publisher}{Springer Berlin Heidelberg},
  \bibinfo{address}{Berlin, Heidelberg}, \bibinfo{year}{2005}).

\bibitem{saleh_anderson_2017}
\bibinfo{author}{Saleh, M.~F.}, \bibinfo{author}{Conti, C.} \&
  \bibinfo{author}{Biancalana, F.}
\newblock \bibinfo{journal}{\bibinfo{title}{Anderson localisation and
  optical-event horizons in rogue-soliton generation}}.
\newblock {\emph{\JournalTitle{Opt. Express}}} \textbf{\bibinfo{volume}{25}},
  \bibinfo{pages}{5457}, \doiprefix\url{10.1364/OE.25.005457}
  (\bibinfo{year}{2017}).

\bibitem{ricard_effects_2024}
\bibinfo{author}{Ricard, G.}, \bibinfo{author}{Novkoski, F.} \&
  \bibinfo{author}{Falcon, E.}
\newblock \bibinfo{journal}{\bibinfo{title}{Effects of nonlinearity on
  {Anderson} localization of surface gravity waves}}.
\newblock {\emph{\JournalTitle{Nat Commun}}} \textbf{\bibinfo{volume}{15}},
  \bibinfo{pages}{5726}, \doiprefix\url{10.1038/s41467-024-49575-5}
  (\bibinfo{year}{2024}).

\bibitem{noauthor_volatility_2024}
\bibinfo{title}{Volatility {Index}® {Methodology}: {Cboe} {Volatility}
  {Index}®} (\bibinfo{year}{2024}).

\bibitem{seegopaul_vstoxx_2024}
\bibinfo{author}{Seegopaul, H.} \& \bibinfo{author}{Shuttlewood, T.}
\newblock \bibinfo{title}{{VSTOXX} 101: {Understanding} {Europe}’s volatility
  benchmark} (\bibinfo{year}{2024}).

\bibitem{akhmediev_editorial_2010}
\bibinfo{author}{Akhmediev, N.} \& \bibinfo{author}{Pelinovsky, E.}
\newblock \bibinfo{journal}{\bibinfo{title}{Editorial – {Introductory}
  remarks on “{Discussion} \& {Debate}: {Rogue} {Waves} – {Towards} a
  {Unifying} {Concept}?”}}.
\newblock {\emph{\JournalTitle{Eur. Phys. J. Spec. Top.}}}
  \textbf{\bibinfo{volume}{185}}, \bibinfo{pages}{1--4},
  \doiprefix\url{10.1140/epjst/e2010-01233-0} (\bibinfo{year}{2010}).

\bibitem{dysthe_oceanic_2008}
\bibinfo{author}{Dysthe, K.}, \bibinfo{author}{Krogstad, H.~E.} \&
  \bibinfo{author}{Müller, P.}
\newblock \bibinfo{journal}{\bibinfo{title}{Oceanic {Rogue} {Waves}}}.
\newblock {\emph{\JournalTitle{Annu. Rev. Fluid Mech.}}}
  \textbf{\bibinfo{volume}{40}}, \bibinfo{pages}{287--310},
  \doiprefix\url{10.1146/annurev.fluid.40.111406.102203}
  (\bibinfo{year}{2008}).

\bibitem{cattrell_can_2018}
\bibinfo{author}{Cattrell, A.~D.}, \bibinfo{author}{Srokosz, M.},
  \bibinfo{author}{Moat, B.~I.} \& \bibinfo{author}{Marsh, R.}
\newblock \bibinfo{journal}{\bibinfo{title}{Can {Rogue} {Waves} {Be}
  {Predicted} {Using} {Characteristic} {Wave} {Parameters}?}}
\newblock {\emph{\JournalTitle{Journal of Geophysical Research: Oceans}}}
  \textbf{\bibinfo{volume}{123}}, \bibinfo{pages}{5624--5636},
  \doiprefix\url{10.1029/2018JC013958} (\bibinfo{year}{2018}).
\newblock \bibinfo{note}{\_eprint:
  https://agupubs.onlinelibrary.wiley.com/doi/pdf/10.1029/2018JC013958}.

\bibitem{mafi_transverse_2015}
\bibinfo{author}{Mafi, A.}
\newblock \bibinfo{journal}{\bibinfo{title}{Transverse {Anderson} localization
  of light: a tutorial}}.
\newblock {\emph{\JournalTitle{Adv. Opt. Photon., AOP}}}
  \textbf{\bibinfo{volume}{7}}, \bibinfo{pages}{459--515},
  \doiprefix\url{10.1364/AOP.7.000459} (\bibinfo{year}{2015}).

\bibitem{degiannakis_forecasting_2018}
\bibinfo{author}{Degiannakis, S.}, \bibinfo{author}{Filis, G.} \&
  \bibinfo{author}{Hassani, H.}
\newblock \bibinfo{journal}{\bibinfo{title}{Forecasting global stock market
  implied volatility indices}}.
\newblock {\emph{\JournalTitle{Journal of Empirical Finance}}}
  \textbf{\bibinfo{volume}{46}}, \bibinfo{pages}{111--129},
  \doiprefix\url{10.1016/j.jempfin.2017.12.008} (\bibinfo{year}{2018}).

\end{thebibliography}

\clearpage

\renewcommand{\figurename}{Supplementary Figure}
\renewcommand{\tablename}{Supplementary Table}

\section*{Supplementary results}

\setcounter{figure}{0}
\setcounter{table}{0}

\begin{figure*}[ht]
\centering
   \includegraphics[width=1\linewidth]{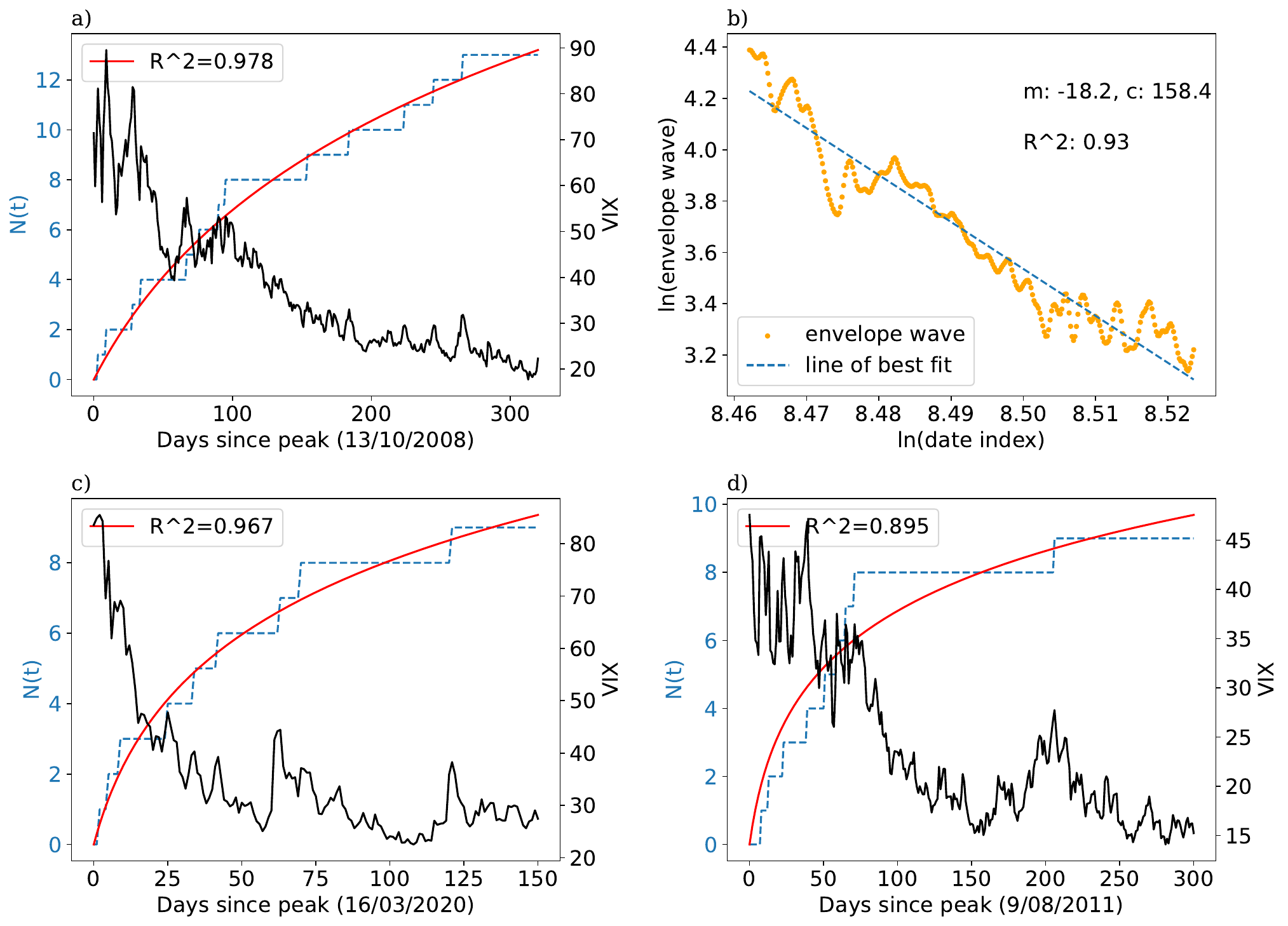}
 
    \caption{a) Left: N(t), cumulative number of times the SWH for the raw VIX data is exceeded following the 2008 peak in the VIX, with a red line displaying a fit to an Omori-like power law seen in Lillo et. al 2003~\cite{lillo_power-law_2003}. Right: VIX highs (unprocessed). b) A log-log plot of the VIX envelope wave vs the date index alongside a linear fit. c) Same as panel a) for the days following the 2020 peak. d) same as panel a) for the days following the 2011 peak. }\label{supfig:Omori}
    
\end{figure*}

\begin{figure*}[ht]
\centering
   \includegraphics[width=1\linewidth]{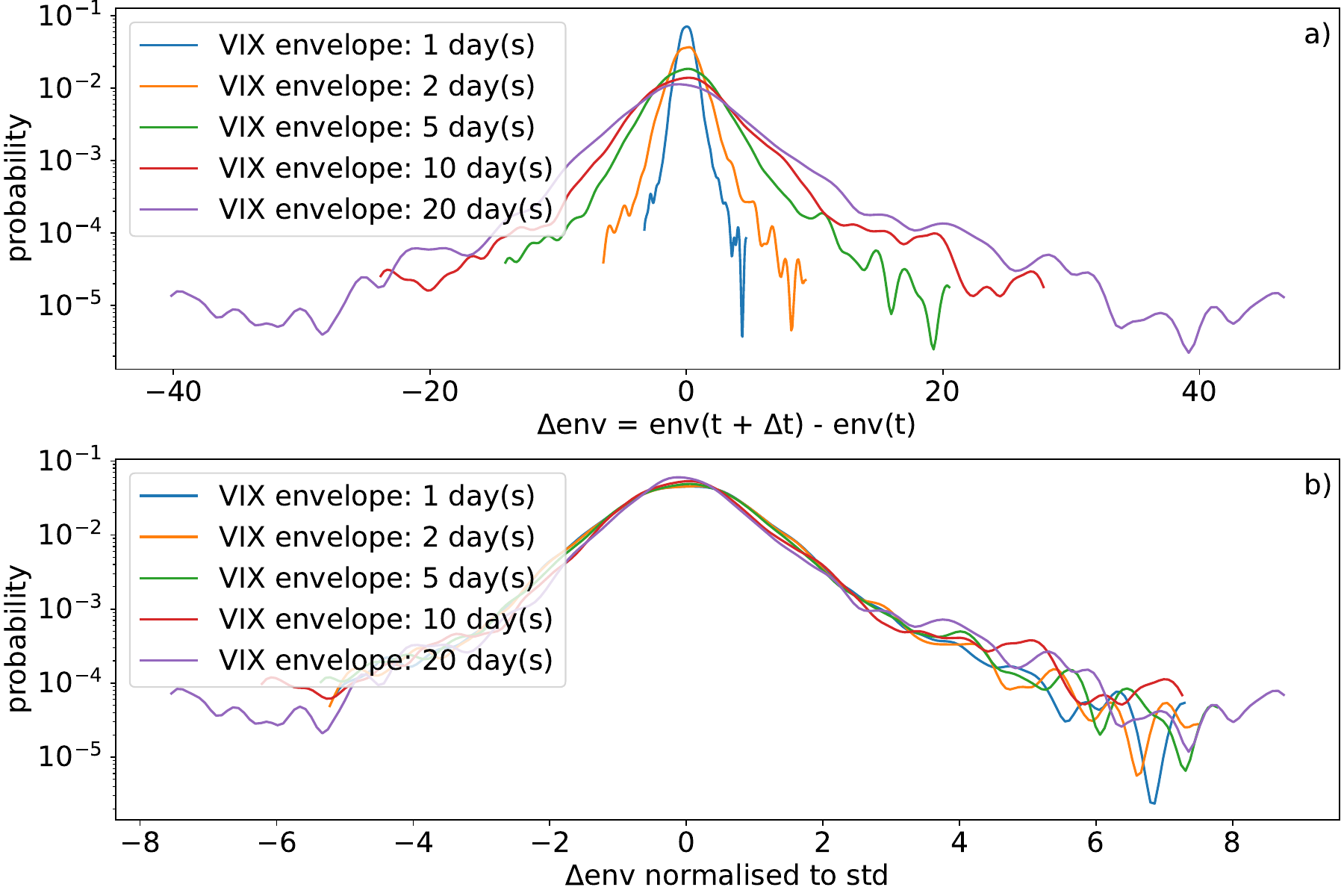}
 
    \caption{a) The estimated probability distribution of differences in VIX envelope values, when the values of the wave envelope are compared for an increasing number of days. Different colours correspond to an increasing number of days, with the distribution spreading out for increasing time scales; note that this is not clearly visible for the raw VIX data. b) The same results with the differences in the envelope wave normalised to their standard deviation, such that the x axis displays $(\Delta \mathrm{env} - \mu)/\sigma)$. Here, we see no change over different time scales. The time scales here are very different from those in the literature where financial returns have been compared to turbulent flow, where a consistent observation is that the tails of the distribution should ``lose weight'' with an increasing time delay~\cite{ghashghaie_turbulent_1996, hutt_synergetics_2020}. This is because, for a slowly-varying envelope, we are above the time differences where scaling of moments appears. The tails of these distributions nevertheless display behaviour which is far from normal, implying long memory seen at shorter timescales persists in the VIX envelope wave.}\label{supfig:turbulent}
\end{figure*}

\begin{figure*}[ht]
\centering
   \includegraphics[width=\linewidth]{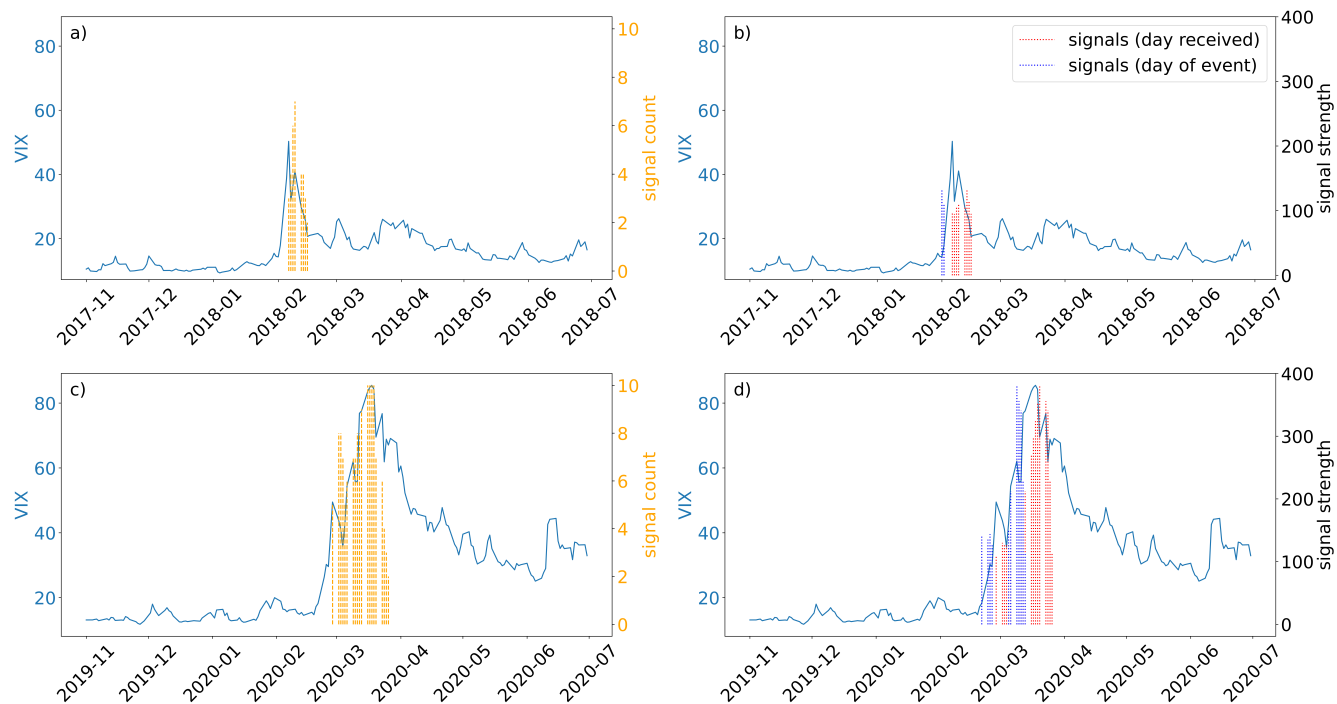}   
    \caption{Comparison of signal counts (number of times the maximum eigenvalue gradient exceeded 50 in the last ten days) and signal strengths (maximum eigenvalue gradient in the last ten days) for the VIX. a) 2018 VIX (blue, left axis) and signal counts (orange dashed, right axis) on the day the signal is received. b) 2018 VIX (blue, left axis) and signals, both on the day of event (dark blue, dotted, right axis) and the day the signal is received (red, dotted, right axis). The blue signals are concentrated on two single days in 2018, suggesting that the eigenvalue gradient isn't continuing to grow -- its maximum values are concentrated on the same dates even with further data. The repeated red lines come from the fact the strong values of the eigenvalue gradient on these two days are being picked up repeatedly. The signal count drops quickly after reaching its maximum. c) 2020 VIX (blue, left axis) and signal counts (orange dashed, left axis) on the day the signal is received. d) 2020 VIX (blue, left axis) and signals, both on the day of event (dark blue, dotted, right axis) and the day the signal is received (red, dotted, right axis). Here, the maximum eigenvalue gradient appears to move forward in time, showing an escalation. }\label{supfig:vix2018}
\end{figure*}

\begin{figure*}[ht]
\centering
   \includegraphics[width=\linewidth]{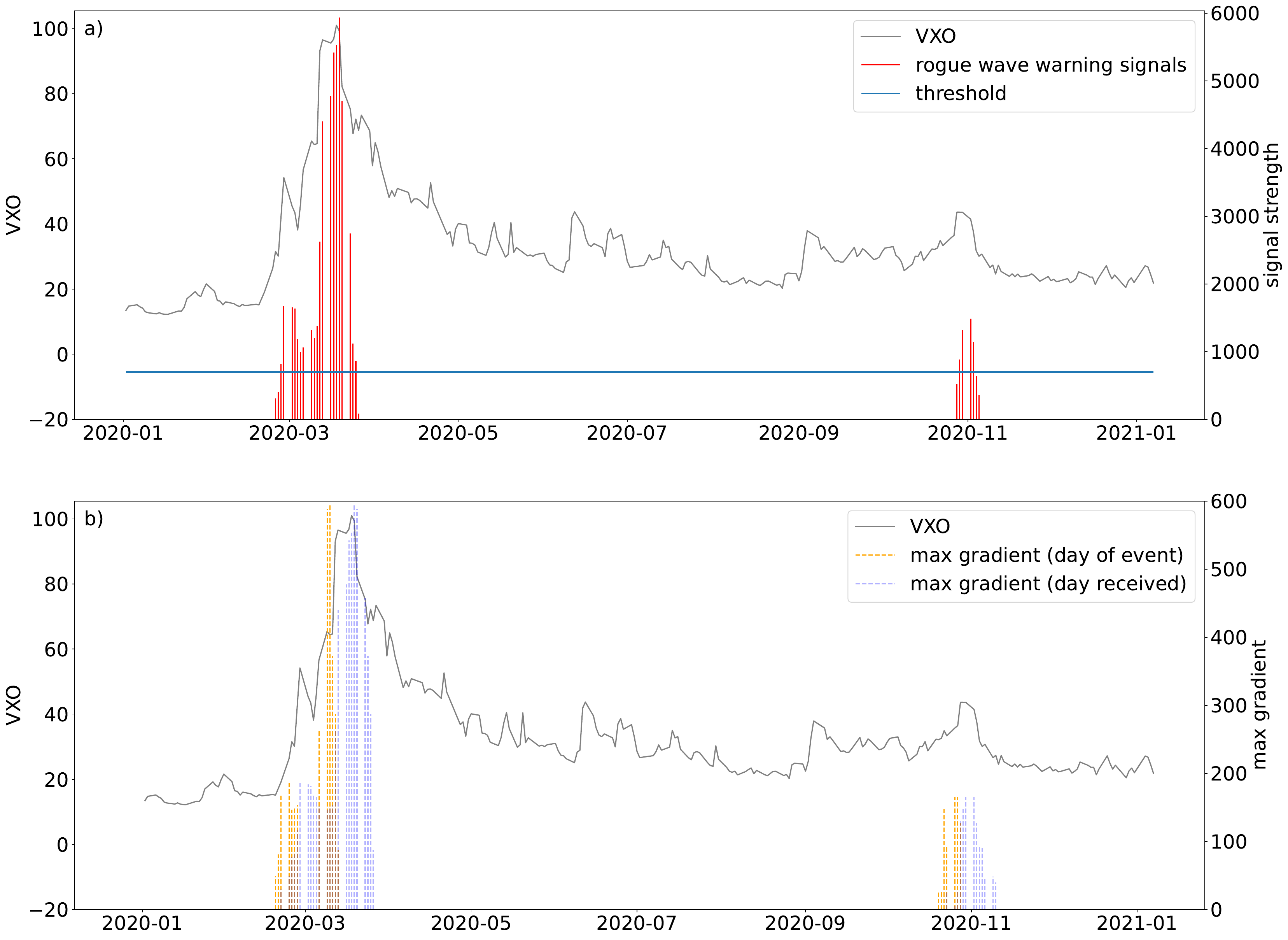}   
    \caption{Comparison of the rogue wave warning signals for the VXO with the maximum eigenvalue gradient on the day of event vs the day the signal is received. Panel a) displays the VXO (grey), rogue wave warning signals (red) and threshold (blue). b) VXO (grey), the maximum eigenvalue gradient on the day of the event (orange dotted) and the day it is recorded (blue dotted). The warnings quickly drop below the threshold after peaking implying a short term event in November. Comparatively, signals diminish but do not drop below the threshold in March before reversing the trend and continuing to grow. }\label{supfig:vxoreal2020}
\end{figure*}

\begin{landscape}
\begin{table}[]
\begin{tabular}{lllllllllllll}
 &
  date &
  first signals &
  days to peak &
  days to peak (low) &
  peaks &
  peaks (low) &
  false  &
  late &
  low  &
  late (low) &
  no signal &
  no signal (low)  \\
0  & 1991-01-14 & --  & 1716.0 & 0.0   & --  & 1.0 & -- & --  & -- & --  & --  & 1.0 \\
1  & 1997-10-28 & --  & 0.0    & 0.0   & 1.0 & 1.0 & -- & --  & -- & --  & 1.0 & 1.0 \\
2  & 1997-10-30 & 1.0 & 210.0  & 210.0 & --  & --  & -- & 1.0 & -- & --  & --  & --  \\
3  & 1998-09-02 & --  & 0.0    & 0.0   & 1.0 & 1.0 & -- & --  & -- & --  & 1.0 & 1.0 \\
4  & 2001-09-17 & 1.0 & 2.0    & 2.0   & --  & --  & -- & --  & -- & --  & --  & --  \\
5  & 2001-09-19 & --  & 0.0    & 0.0   & 1.0 & 1.0 & -- & --  & -- & --  & --  & --  \\
6  & 2002-07-24 & 1.0 & 52.0   & 52.0  & --  & --  & -- & --  & -- & --  & --  & --  \\
7  & 2002-10-07 & --  & 0.0    & 0.0   & 1.0 & 1.0 & -- & --  & -- & --  & --  & --  \\
8  & 2008-09-18 & 1.0 & 18.0   & 18.0  & --  & --  & -- & --  & -- & --  & --  & --  \\
9  & 2008-09-24 & 1.0 & 14.0   & 14.0  & --  & --  & -- & --  & -- & --  & --  & --  \\
10 & 2008-10-02 & 1.0 & 8.0    & 8.0   & --  & --  & -- & --  & -- & --  & --  & --  \\
11 & 2008-10-14 & --  & 0.0    & 0.0   & 1.0 & 1.0 & -- & --  & -- & --  & --  & --  \\
12 & 2010-05-10 & 1.0 & 8.0    & 8.0   & --  & --  & -- & --  & -- & --  & --  & --  \\
13 & 2010-05-13 & 1.0 & 5.0    & 5.0   & --  & --  & -- & --  & -- & --  & --  & --  \\
14 & 2010-05-19 & 1.0 & 1.0    & 1.0   & --  & --  & -- & --  & -- & --  & --  & --  \\
15 & 2010-05-20 & --  & 0.0    & 0.0   & 1.0 & 1.0 & -- & --  & -- & --  & --  & --  \\
16 & 2011-08-08 & 1.0 & 1.0    & 1.0   & --  & --  & -- & --  & -- & --  & --  & --  \\
17 & 2011-08-09 & --  & 0.0    & 0.0   & 1.0 & 1.0 & -- & --  & -- & --  & --  & --  \\
18 & 2015-08-25 & 1.0 & 1.0    & 1.0   & --  & --  & -- & --  & -- & --  & --  & --  \\
19 & 2015-08-26 & --  & 0.0    & 0.0   & 1.0 & 1.0 & -- & --  & -- & --  & --  & --  \\
20 & 2018-02-06 & 1.0 & 0.0    & 0.0   & 1.0 & 1.0 & -- & 1.0 & -- & --  & 1.0 & 1.0 \\
21 & 2018-12-26 & --  & 306.0  & 0.0   & --  & 1.0 & -- & --  & -- & --  & --  & 1.0 \\
22 & 2020-02-28 & 1.0 & 11.0   & 11.0  & --  & --  & -- & --  & -- & --  & --  & --  \\
23 & 2020-03-16 & --  & 0.0    & 0.0   & 1.0 & 1.0 & -- & --  & -- & --  & --  & --  \\
24 & 2021-12-06 & 1.0 & 212.0  & 212.0 & --  & --  & -- & --  & -- & --  & --  & --  \\
25 & 2022-01-27 & 1.0 & 176.0  & 176.0 & --  & --  & -- & --  & -- & --  & --  & --  \\
26 & 2022-02-01 & 1.0 & 173.0  & 173.0 & --  & --  & -- & --  & -- & --  & --  & --  \\
27 & 2022-10-04 & --  & 0.0    & 0.0   & 1.0 & 1.0 & -- & --  & -- & --  & --  & --  \\
28 & 2024-08-05 & 1.0 & 174.0  & 0.0   & --  & 1.0 & -- & --  & -- & 1.0 & --  & 1.0 \\
29 & 2025-03-11 & 1.0 & 20.0   & 20.0  & --  & --  & -- & --  & -- & --  & --  & --  \\
30 & 2025-04-07 & 1.0 & 1.0    & 1.0   & --  & --  & -- & --  & -- & --  & --  & --  \\
31 & 2025-04-08 & --  & 0.0    & 0.0   & 1.0 & 1.0 & -- & --  & -- & --  & --  & -- 
\end{tabular}
\caption{The first rogue wave warning signals in each series of signals for the VIX, and all peaks, marked by their dates. Here, low denotes when the threshold of 2$\times$SWH was used instead of 2.5$\times$SWH. See the methods section for details on how signal categories were assigned.}
\label{suptable:vix}
\end{table}
\end{landscape}

\begin{landscape}
\begin{table}[]
\begin{tabular}{lllllllllllll}
 &
  date &
  first signals &
  days to peak &
  days to peak (low) &
  peaks &
  peaks (low) &
  false &
  late  &
  low  &
  late (low) &
  no signal &
  no signal (low)  \\
0  & 1987-10-19 & 1.0 & 2.0    & 2.0    & --  & --  & --  & --  & --  & --  & --  & --  \\
1  & 1987-10-21 & --  & 0.0    & 0.0    & 1.0 & 1.0 & --  & --  & --  & --  & --  & --  \\
2  & 1989-10-18 & 1.0 & 2261.0 & 314.0  & --  & --  & 1.0 & --  & --  & --  & --  & --  \\
3  & 1990-08-08 & 1.0 & 2058.0 & 111.0  & --  & --  & --  & --  & 1.0 & --  & --  & --  \\
4  & 1991-01-16 & --  & 1947.0 & 0.0    & --  & 1.0 & --  & --  & --  & --  & --  & --  \\
5  & 1991-06-06 & 1.0 & 1850.0 & 1626.0 & --  & --  & 1.0 & --  & --  & --  & --  & --  \\
6  & 1997-10-30 & 1.0 & 234.0  & 10.0   & --  & --  & --  & --  & --  & --  & --  & --  \\
7  & 1997-11-03 & 1.0 & 232.0  & 8.0    & --  & --  & --  & --  & --  & --  & --  & --  \\
8  & 1997-11-13 & --  & 224.0  & 0.0    & --  & 1.0 & --  & --  & --  & --  & --  & --  \\
9  & 1998-09-01 & 1.0 & 25.0   & 25.0   & --  & --  & --  & --  & --  & --  & --  & --  \\
10 & 1998-10-07 & --  & 0.0    & 0.0    & 1.0 & 1.0 & --  & --  & --  & --  & --  & --  \\
11 & 2001-09-18 & 1.0 & 2.0    & 2.0    & --  & --  & --  & --  & --  & --  & --  & --  \\
12 & 2001-09-20 & --  & 0.0    & 0.0    & 1.0 & 1.0 & --  & --  & --  & --  & --  & --  \\
13 & 2002-07-16 & 1.0 & 61.0   & 61.0   & --  & --  & --  & --  & --  & --  & --  & --  \\
14 & 2002-07-23 & 1.0 & 56.0   & 56.0   & --  & --  & --  & --  & --  & --  & --  & --  \\
15 & 2002-07-25 & 1.0 & 54.0   & 54.0   & --  & --  & --  & --  & --  & --  & --  & --  \\
16 & 2002-10-10 & --  & 0.0    & 0.0    & 1.0 & 1.0 & --  & --  & --  & --  & --  & --  \\
17 & 2008-09-19 & 1.0 & 14.0   & 14.0   & --  & --  & --  & --  & --  & --  & --  & --  \\
18 & 2008-10-01 & 1.0 & 6.0    & 6.0    & --  & --  & --  & --  & --  & --  & --  & --  \\
19 & 2008-10-09 & --  & 0.0    & 0.0    & 1.0 & 1.0 & --  & --  & --  & --  & --  & --  \\
20 & 2010-05-24 & --  & 0.0    & 0.0    & 1.0 & 1.0 & --  & --  & --  & --  & 1.0 & 1.0 \\
21 & 2010-05-25 & 1.0 & 342.0  & 342.0  & --  & --  & --  & 1.0 & --  & --  & --  & --  \\
22 & 2011-08-09 & 1.0 & 37.0   & 37.0   & --  & --  & --  & --  & --  & --  & --  & --  \\
23 & 2011-09-30 & --  & 0.0    & 0.0    & 1.0 & 1.0 & --  & --  & --  & --  & --  & --  \\
24 & 2015-08-26 & 1.0 & 1145.0 & 2.0    & --  & --  & --  & --  & 1.0 & --  & --  & --  \\
25 & 2015-08-28 & --  & 1143.0 & 0.0    & --  & 1.0 & --  & --  & --  & --  & --  & --  \\
26 & 2018-02-07 & --  & 528.0  & 0.0    & --  & 1.0 & --  & --  & --  & --  & --  & 1.0 \\
27 & 2018-12-24 & --  & 307.0  & 0.0    & --  & 1.0 & --  & --  & --  & --  & --  & 1.0 \\
28 & 2018-12-27 & 1.0 & 305.0  & 305.0  & --  & --  & --  & --  & --  & 1.0 & --  & --  \\
29 & 2020-02-27 & 1.0 & 12.0   & 12.0   & --  & --  & --  & --  & --  & --  & --  & --  \\
30 & 2020-03-16 & --  & 0.0    & 0.0    & 1.0 & 1.0 & --  & --  & --  & --  & --  & --  \\
31 & 2020-10-29 & 1.0 & --     & --     & --  & --  & 1.0 & --  & --  & --  & --  & -- 
\end{tabular}
\caption{The first rogue wave warning signals in each series of signals for the VXO, and all peaks, marked by their dates. Here, low denotes when the threshold of 2$\times$SWH was used instead of 2.5$\times$SWH. See the methods section for details on how categories were assigned.}
\label{suptable:vxo}
\end{table}
\end{landscape}

\begin{landscape}
\begin{table}[]
\begin{tabular}{lllllllllllll}
 &
  date &
  first signals &
  days to peak &
  days to peak (low) &
  peaks &
  peaks (low) &
  false &
  late &
  low  &
  late (low) &
  no signal &
  no signal (low)  \\
0  & 2001-09-11 & 1.0 & 6.0    & 6.0    & --  & --  & --  & --  & --  & -- & --  & --  \\
1  & 2001-09-19 & --  & 0.0    & 0.0    & 1.0 & 1.0 & --  & --  & --  & -- & --  & --  \\
2  & 2002-07-17 & 1.0 & 55.0   & 55.0   & --  & --  & --  & --  & --  & -- & --  & --  \\
3  & 2002-09-30 & 1.0 & 2.0    & 2.0    & --  & --  & --  & --  & --  & -- & --  & --  \\
4  & 2002-10-02 & --  & 0.0    & 0.0    & 1.0 & 1.0 & --  & --  & --  & -- & --  & --  \\
5  & 2008-01-24 & 1.0 & 185.0  & 185.0  & --  & --  & --  & --  & --  & -- & --  & --  \\
6  & 2008-09-19 & 1.0 & 17.0   & 17.0   & --  & --  & --  & --  & --  & -- & --  & --  \\
7  & 2008-10-01 & 1.0 & 9.0    & 9.0    & --  & --  & --  & --  & --  & -- & --  & --  \\
8  & 2008-10-06 & 1.0 & 6.0    & 6.0    & --  & --  & --  & --  & --  & -- & --  & --  \\
9  & 2008-10-14 & --  & 0.0    & 0.0    & 1.0 & 1.0 & --  & --  & --  & -- & --  & --  \\
10 & 2008-10-29 & 1.0 & 392.0  & 392.0  & --  & --  & 1.0 & --  & --  & -- & --  & --  \\
11 & 2010-05-07 & 1.0 & 9.0    & 9.0    & --  & --  & --  & --  & --  & -- & --  & --  \\
12 & 2010-05-20 & --  & 0.0    & 0.0    & 1.0 & 1.0 & --  & --  & --  & -- & --  & --  \\
13 & 2010-05-21 & 1.0 & 337.0  & 337.0  & --  & --  & --  & 1.0 & --  & -- & --  & --  \\
14 & 2011-08-08 & 1.0 & 25.0   & 25.0   & --  & --  & --  & --  & --  & -- & --  & --  \\
15 & 2011-09-12 & --  & 0.0    & 0.0    & 1.0 & 1.0 & --  & --  & --  & -- & --  & --  \\
16 & 2011-09-13 & 1.0 & 1210.0 & 1210.0 & --  & --  & --  & 1.0 & --  & -- & --  & --  \\
17 & 2016-06-21 & --  & 0.0    & 0.0    & 1.0 & 1.0 & --  & --  & --  & -- & 1.0 & 1.0 \\
18 & 2018-02-08 & 1.0 & 531.0  & 531.0  & --  & --  & 1.0 & --  & --  & -- & --  & --  \\
19 & 2020-02-27 & 1.0 & 12.0   & 12.0   & --  & --  & --  & --  & --  & -- & --  & --  \\
20 & 2020-03-16 & --  & 0.0    & 0.0    & 1.0 & 1.0 & --  & --  & --  & -- & --  & --  \\
21 & 2022-02-28 & 1.0 & 4.0    & 4.0    & --  & --  & --  & --  & --  & -- & --  & --  \\
22 & 2022-03-02 & 1.0 & 2.0    & 2.0    & --  & --  & --  & --  & --  & -- & --  & --  \\
23 & 2022-03-04 & --  & 0.0    & 0.0    & 1.0 & 1.0 & --  & --  & --  & -- & --  & --  \\
24 & 2025-04-07 & 1.0 & --     & 2.0    & --  & --  & --  & --  & 1.0 & -- & --  & --  \\
25 & 2025-04-09 & --  & --     & 0.0    & --  & 1.0 & --  & --  & --  & -- & --  & -- 
\end{tabular}
\caption{The first rogue wave warning signals in each series of signals for the VSTOXX, and all peaks, marked by their dates. Here, low denotes when the threshold of 2$\times$SWH was used instead of 2.5$\times$SWH. See the methods section for details on how categories were assigned.}
\label{suptable:vstoxx}
\end{table}
\end{landscape}

\end{document}